\UseRawInputEncoding
\documentclass[aps,bibliography]{revtex4-1}
\usepackage{epsfig}
\usepackage{graphicx}
\usepackage{soul,amsmath,amssymb}
\usepackage[breaklinks=true,colorlinks]{hyperref}

\newcommand{\bn}{\begin{eqnarray}}
\newcommand{\en}{\end{eqnarray}}
\newcommand{\eml}{\end{multline}}
\newcommand{\bml}{\begin{multline}}

\begin{document}

\title {Work extraction from single-mode thermal noise by measurements: How important is information?}
\author{Avijit Misra}
\email{avijit.misra@weizmann.ac.il}
\affiliation{AMOS and Department of Chemical and Biological Physics,
Weizmann Institute of Science, Rehovot 7610001, Israel}
\affiliation{International Center of Quantum Artificial Intelligence for Science and Technology (QuArtist)
and Department of Physics, Shanghai University, 200444 Shanghai, China}
\author{Tomas Opatrn\'y}
\email{tomas.opatrny@upol.cz}
\affiliation{Department of Optics, Faculty of Science, Palack\'y University, 17. listopadu 50, 77146 Olomouc, Czech Republic}

\author{Gershon Kurizki}
\email{gershon.kurizki@weizmann.ac.il}
\affiliation{AMOS and Department of Chemical and Biological Physics,
Weizmann Institute of Science, Rehovot 7610001, Israel}
 
\begin{abstract}
Our goal in this article is  to  elucidate the rapport of work and information in the context of a minimal quantum mechanical setup:   A converter of heat input to work output,  the  input consisting of a single oscillator mode prepared in a hot thermal state  along with few much colder oscillator modes.   We wish to achieve heat to work conversion in the setup while avoiding the use of a working substance (medium)  or macroscopic heat baths. The core issues we consider, taking account of the quantum nature of the setup, are: (i) How and to what extent can information act as work resource or, conversely, be redundant for work extraction? (ii)  What is the optimal way of extracting work via information acquired by measurements? (iii) What is the bearing of information on the efficiency-power tradeoff achievable in such setups? We compare the efficiency of work extraction and the limitations of power in our minimal setup by unitary (reversible) manipulations and  by  different, generic,  measurement strategies of the hot and cold  modes. For each strategy the rapport of work and information extraction is found and the cost of information erasure is allowed for. The possibilities of work extraction without information acquisition, via non-selective measurements, are also analyzed. Overall, we present, by  generalizing a method  based on optimized homodyning that we have recently proposed, the following insight: extraction of work by observation and feedforward (WOF) that only measures a {\it small fraction} of the input,  is clearly advantageous to the conceivable alternatives. Our results may become a basis of a practical strategy of converting thermal noise to useful work in optical setups, such as coherent amplifiers of thermal light, as well as in their optomechanical and photovoltaic counterparts.
\end{abstract}

\date{\today}

 \maketitle
 
\section{Introduction}
\label{Intro}

\noindent Thermal noise, i.e. maximal-entropy fluctuation at a given temperature, is a ubiquitous source of propagating  energy, ranging from sunlight and cosmic rays to acoustic (e.g. seismic) energy. Since the invention of the steam engine technology has aimed at harnessing  thermal noise (heat)  to the performance of useful work. The definition of work in the literature is elusive, but may be loosely phrased as the most ordered energy, or, more formally, as energy exchange with the least (ideally zero) entropy exchange. The question to be posed is: what is the {\it most efficient way} of accomplishing such heat-to-work conversion? Not less importantly: what is the {\it fastest way} of converting heat to work, thereby attaining  the maximal rate of work production, alias {\it maximal power}?

 The conversion of heat to work consists in lowering the entropy from its highest to its lowest value at a given energy within the constraints of the first and second laws of thermodynamics \cite{Schwabl_BOOK}. Such conversion is a central theme that quantum thermodynamics (QTD) has inherited from its classical predecessor \cite{Kurizkibook,gemmerbook,scovil1959three,kosloff2014quantum,Kosloff_2013,Gelbwaser_2013_a,David15}, the difference being that  QTD accounts for possible effects of  coherence and entanglement in heat engine (HE) designs \cite{Scully_2003,dillenschneider2009energetics,Liberato_2011,Huang_2012,abah14,Li_2014,abah2012single,hardal2015superradiant,Dag_2016,Niedenzu16,Turkpence_2016,Correa_2014,niedenzu18quantum,scully2011quantum,uzdin2015quantum,Ghosh2018,PhysRevLett.119.170602,Cooperative,Collectively,QTM2,Klatzow,kla17,FTE4,espositoPRL10,Deffner,CarnotPRE,RF2}. A conceptual alternative to HE has been provided by information engines (IE)  originating  from Maxwell-demon \cite{Maxwell} and Szilard engines \cite{Szilard1929,LandauerIBM61,BennettIJTP82,parrondo2015thermodynamics}, which exploit information acquired by measurements as a resource complementary to heat. Both HE and IE have merits but also basic limitations, the central one being power-efficiency tradeoff: In HE it is inevitable for power to diminish near the point of maximum efficiency,  which is  always bounded by the Carnot limit in accordance with the second law \cite{Kurizkibook,David15,scovil1959three,Schwabl_BOOK,Gelbwaser_2013_a,gemmerbook,scovil1959three,kosloff2014quantum,Kosloff_2013,FTE4,espositoPRL10,Deffner,CarnotPRE,RF2}. Commonly, IE have been based on binary measurements of discrete variables, whose energetic price  yields efficiency bounds  well below unity \cite{Vid2016,SagawaPRL08,Elouard18,Elouard17}. On the other hand, the duration of work extraction from IE is not intrinsically related to the efficiency, so that their power-efficiency tradeoff may be in principle more favorable than in HE, as indeed is shown here.
 
An HE concentrates the energy of a  heat bath, which extends over macroscopic numbers of modes,  into a  single working mode. The macroscopic nature of the heat baths entails the description of HE as dissipative, open systems, typically treated by master equations \cite{Kurizkibook,gemmerbook,scovil1959three,kosloff2014quantum,Kosloff_2013,Gelbwaser_2013_a,David15}. A major part of our motivation here is to inquire: Is it  truly necessary to resort to macroscopic heat baths  and open-system  treatment  of HE operation in the quantum domain or  can we describe them in much simpler terms as closed, finite quantum systems? We opt for the latter, simple and tractable description of HE as {\it few-mode systems driven by classical fields or measured by detectors.} 

Our goal is  to  elucidate the rapport of work and information in the context of a minimal quantum mechanical setup:   A converter of heat input to work output,  the  input consisting of a single oscillator mode prepared in a hot thermal state  along with $N-1$ much colder oscillator modes, all initially at the same frequency.   We wish to achieve heat to work conversion in the setup while avoiding the use of a working substance (medium)  or macroscopic heat baths.  Grosso modo, there are two alternative ways of achieving this goal: (i) either we {\it unitarily} manipulate the mode frequencies and their  contact with each other, or (ii) we measure the input, and exploit the measured results  as feedforward for work extraction. The latter option concerns the acquisition of information by measurements, its cost and utilization for our purpose. The core issues we consider, taking account of the quantum nature of the setup, are: 1) How and to what extent can information act as work resource or, conversely, be redundant for work extraction? 2)  What is the optimal way of extracting work via information acquired by measurements? 3) What is the bearing of information on the efficiency-power tradeoff achievable in such setups? 
To address these questions we first restate nonequilibrium heat to work conversion in terms of ergotropy (non-passivity, Sec. \ref{secII}) \cite{Allahverdyan_2004_a,Pusz_1978,Gelbwaser_2013_b,David15,niedenzu18quantum,Alicki_1979,Pas6,Gelbwaser_2015_b,Gelbwaser_2013_b,David15,PRE2014,Ghosh2017,Ghosh2018,PRE2014,Gelbwaser_2013_a,Kurizkibook}. We then  compare the efficiency of work extraction and the limitations of power in our minimal setup by unitary (reversible) manipulations (Sec. \ref{secIII}) and  by  different, generic,  measurement strategies of the hot and cold  modes (Sec. \ref{secIV}-\ref{secV}), finding for each strategy the rapport of work and information extraction and allowing for the cost of information erasure. The possibilities of work extraction without information acquisition, via non-selective measurements, are analyzed (Sec. \ref{secVI}) The findings are summarized  in  the Conclusions (Sec. \ref{secVII}). Overall, we present, by  generalization of a method  based on optimized homodyning we have recently proposed \cite{WOF}, the following insight: extraction of work by observation and feedforward (WOF) that only measures a {\it small fraction} of the input,  is clearly advantageous to the conceivable alternatives. As discussed in the Conclusions, our results may become a basis of a practical strategy of converting thermal noise to useful work in optical setups, such as coherent amplifiers of thermal light, as well as in their optomechanical \cite{Gelbwaser_2015_b} and photovoltaic \cite{Dong2021}  counterparts.

\section{Ergotropy and work extraction from a driven open system}
\label{secII}
\noindent At the outset, we briefly present the key expressions for work and heat extractable from a quantum system driven by an external classical field, without any assumptions on the system state or its dynamics. These expressions will help guide us through the different work extraction processes in Sec. \ref{secIII}-\ref{secVI}.


\noindent Ergotropy is the maximum amount of work extractable for a given Hamiltonian $H$ from a state $\rho$ with mean energy $\langle E\rangle$ by unitary transformations. It is quantified as \cite{Pusz_1978,Gelbwaser_2013_b,David15,niedenzu18quantum,Allahverdyan_2004_a,Pas6,Gelbwaser_2013_a}
\begin{equation}\label{11.eq_app_def_ergotropy}
  \mathcal{W}(\rho,H)\equiv
\mbox{Tr}(\rho H)-\min_U\mbox{Tr}(U\rho U^\dagger H)\geq0,
\end{equation}
where the minimization encompasses all possible unitary transformations $U$. To have $\mathcal{W}(\rho,H)>0$ the state $\rho$ must be non-passive, i.e. correspond to a {\it non-monotonic or anisotropic} distribution of energy eigenvalues. The mean energy $\langle E\rangle$ of such a state $\rho$ can be divided into ergotropy $\mathcal{W}$ and passive energy, i.e. the energy that cannot be extracted as useful work by a unitary operation, which is given by
\begin{equation}
\langle E\rangle-\mathcal{W}=\mbox{Tr}(U_{\rm p}\rho U_{\rm p}^\dagger H)=\mbox{Tr}(\Pi H).
\end{equation}
Here $U_{\rm p}$ is the unitary transformation from state $\rho$ to its (unique) passive counterpart state $\Pi$. This transformation minimizes the second term on the right-hand side of \eqref{11.eq_app_def_ergotropy}.

The change in the mean energy of an evolving system, $\Delta \langle E(t)\rangle$
consists of two qualitatively different contributions:
\begin{equation}\label{11.eq_first_law}
  \Delta \langle E(t)\rangle= W(t)+\mathcal{E}_\mathrm{d}(t).
\end{equation}
Here the term
  \begin{equation}\label{11.eq_def_work}
  W(t)=\int_0^t\mbox{Tr}[\rho(t^\prime)\dot H(t^\prime)]d t^\prime
  \end{equation}
is identified in Alicki's formula \cite{Alicki_1979} as the work that is either extracted or invested by an external drive. It is often considered to be ``classical work'' as opposed to work of ``quantum-coherent'' origin \cite{Seifert-PRL}, although coherence can have classical origin. Hence, this distinction is superfluous \cite{Ghosh2018,QTM2}.

The other term in Eq. (\ref{11.eq_first_law}) has the form
\begin{equation}\label{11.eq_def_DeltaEdiss}
    \mathcal{E}_\mathrm{d}(t)=\int_0^t\mbox{Tr}[\dot\rho(t^\prime)H(t^\prime)]dt^\prime.
  \end{equation}
It is commonly identified with heat exchange \cite{Alicki_1979}. 
Yet, as shown by us \cite{Gelbwaser_2013_b,David15,niedenzu18quantum}, $\mathcal{E}_\mathrm{d}$ may consist of both dissipation (heat exchange) and the exchange of ordered energy, i.e. work, between the system and a bath.

In order to separate these two processes, we decompose $\mathcal{E}_\mathrm{d}(t)$ into passive and non-passive contributions, as
\begin{equation}\label{11.eq_DeltaEdiss_decomposition}
\mathcal{E}_\mathrm{d}(t)=\mathcal{Q}(t)+
\Delta\mathcal{W}_\mathrm{d}(t),
\end{equation}
where
  \begin{equation}\label{11.eq_def_heat}
\mathcal{Q}(t)=
\int_{0}^{t} \mbox{Tr}[\dot{\Pi}(t^\prime)H(t^\prime)] dt^\prime
  \end{equation}
corresponds to a change in the passive state and thus a change in entropy.
Because of its entropy-changing character, we refer  to \eqref{11.eq_def_heat} as {\it heat exchange}:
$\mathcal{Q}(t)$ is the  non-unitary change in passive energy.

The other contribution in \eqref{11.eq_DeltaEdiss_decomposition},
\begin{equation}\label{11.eq_def_DeltaW_diss}
\Delta\mathcal{W}(t)=\int_0^t\mbox{Tr}\Big[\big( \dot\rho(t^\prime)-\dot\Pi(t^\prime)\big)H(t^\prime)\Big] dt^\prime,
\end{equation}
is the dissipative (non-unitary) {\it change in the ergotropy} due to the interaction of the system with the bath.
The state $\rho(t)$ of a system that interacts with a bath and is driven by classical fields typically satisfies the Lindblad master equation \cite{Breuer_BOOK,David15,niedenzu18quantum,Samya}.
If the system state retains its passivity, so that $\Delta\mathcal{W}_\mathrm{d}(t)=0$, the dissipative energy change is then entirely heat exchange, $\mathcal{E}_\mathrm{d}(t)=\mathcal{Q}(t)$.
The ergotropy\index{ergotropy} may increase in a non-unitary fashion due to the interaction of the system with a bath and be subsequently extracted as work via a unitary process. 
Any \textit{unitary} change in the passive energy of a system driven by a time-dependent Hamiltonian results in a change in the extracted work. 

In this article we are concerned with a harmonic oscillator mode for which the driving is so slow that according to Eq. (\ref{11.eq_def_work}) $W(t)=0$. The extractable work is then the ergotropy extractable by displacement (downshift) of the state $\rho$ to the origin \cite{WOF}
\begin{equation}
 \Delta\mathcal{W}_\mathrm{d}= \frac{\hbar \omega}{2} (\langle \hat x\rangle^2+\langle \hat p\rangle^2),
\end{equation}
where $\langle \hat x\rangle$ and $\langle \hat p\rangle$ are the mean values of the position and the momentum, respectively.
\section{Reversible work extraction from thermal mode}
\label{secIII}

\begin{figure}
\includegraphics[width=7cm]{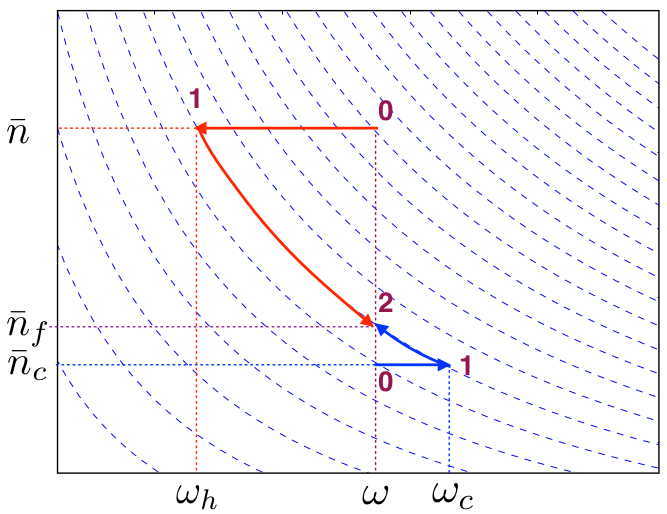}
\caption{\label{schemecarnot}
Schematic consecutive steps $0 \to 1$ and $1 \to 2$ in reversible work extraction for $N=2$ oscillator modes. The  mean quanta number $\bar n$ is plotted versus frequency for the hot mode (red line) and the cold one (blue line). The thin dashed blue curves represent isotherms at various temperatures. The two oscillator modes at the same frequency $\omega$ are initially at
different temperatures corresponding to mean quanta numbers $\bar n$ and
$\bar n_c$. First, the frequencies of
the oscillators are adiabatically changed so that they arrive at the
same isotherm. Then, they are brought into thermal contact and their
frequencies are isothermally restored to the original value. The mean quanta
numbers $\bar n$, $\bar n_c$ and number of modes $N$ determine the mean extracted work as per Eq. (\ref{r-work}).
}
\end{figure}
\begin{figure}
\includegraphics[width=16cm]{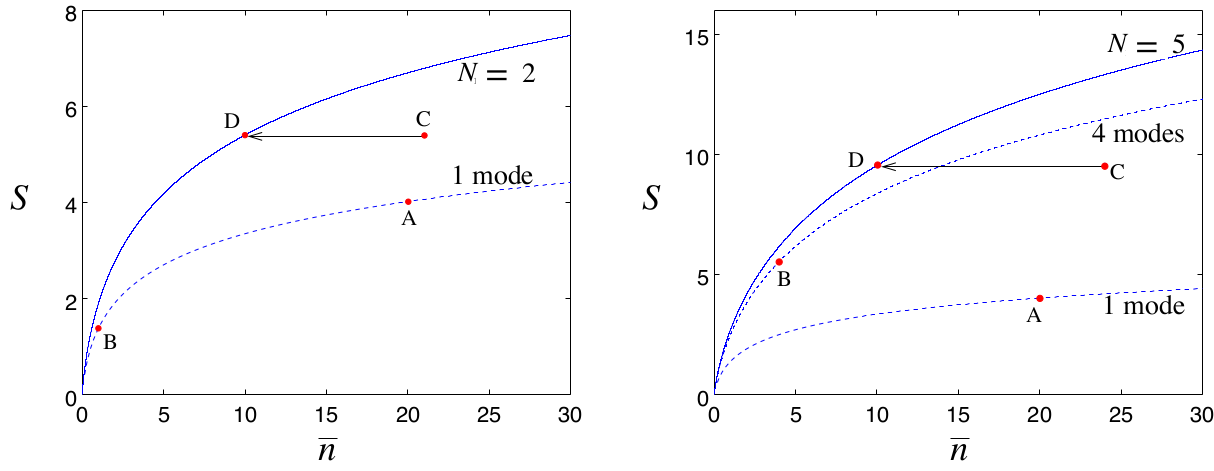}
\caption{\label{f-Carn3}
Extracting work by a reversible process in the entropy (S)-energy ($\bar n$) plane with 2 modes (left panel) and 5 modes (right panel) of harmonic oscillators at the same frequency $\omega$. The system starts with $\bar n=20$ quanta in the hot mode and $\bar n_c=1$ quanta in each of the remaining $N-1$ cold modes. Energy and entropy 
of the initial state of the hot mode correspond to point A, those of the remaining $N-1$ modes to B, and the total $N-$ mode energy and entropy to C. All possible states of the combined  $N$ modes have parameters that are below the full line, which corresponds to thermal states that by definition maximize entropy for a given energy. Work can be extracted by moving from C along the horizontal isentropic line to the equilibrium state D. The horizontal distance between C and D corresponds to work. 
For $N=2$ one can extract the energy of 11.0 quanta out of 20 as work, while for $N=5$ the extractable work is the energy of 13.9 quanta out of 20. 
}
\end{figure}

\noindent As a benchmark for work extraction from a hot oscillator mode, let us consider a cycle involving $N$ modes whose combined entropy does not change throughout the cycle, as it is effected by a sequence of reversible steps:

i) We start from one mode whose thermal field has a mean number of quanta $\bar n$ and  $N-1$ colder modes with $\bar n_c$ quanta on average per mode.
 
ii) A reversible process that does not change the total entropy, redistributes the energy such that the final state is thermal with  $\bar n_f$ quanta in each of the $N$ modes (Fig. \ref{schemecarnot} ). The total number of quanta is not conserved, thus allowing work extraction.
This is the optimal unitary transformation for work extraction, since, all modes being at equal temperature, the final state has zero ergotropy. The final mean number of quanta can be found from the equality between the final(f) and initial(0) entropies (Fig. \ref{f-Carn3}, App. \ref{app:A}):
\begin{eqnarray}
\label{Ebalance}
S_f  (N,\bar n_f) = S_0 (N,\bar n, \bar n_c).
\end{eqnarray}
which yields the extracted work
\begin{eqnarray}
\label{r-work}
W= E_0 - E_f= \hbar \omega[\bar n + (N-1)\bar n_c-N\bar n_f].
\end{eqnarray}

\begin{figure}
\includegraphics[width=10cm]{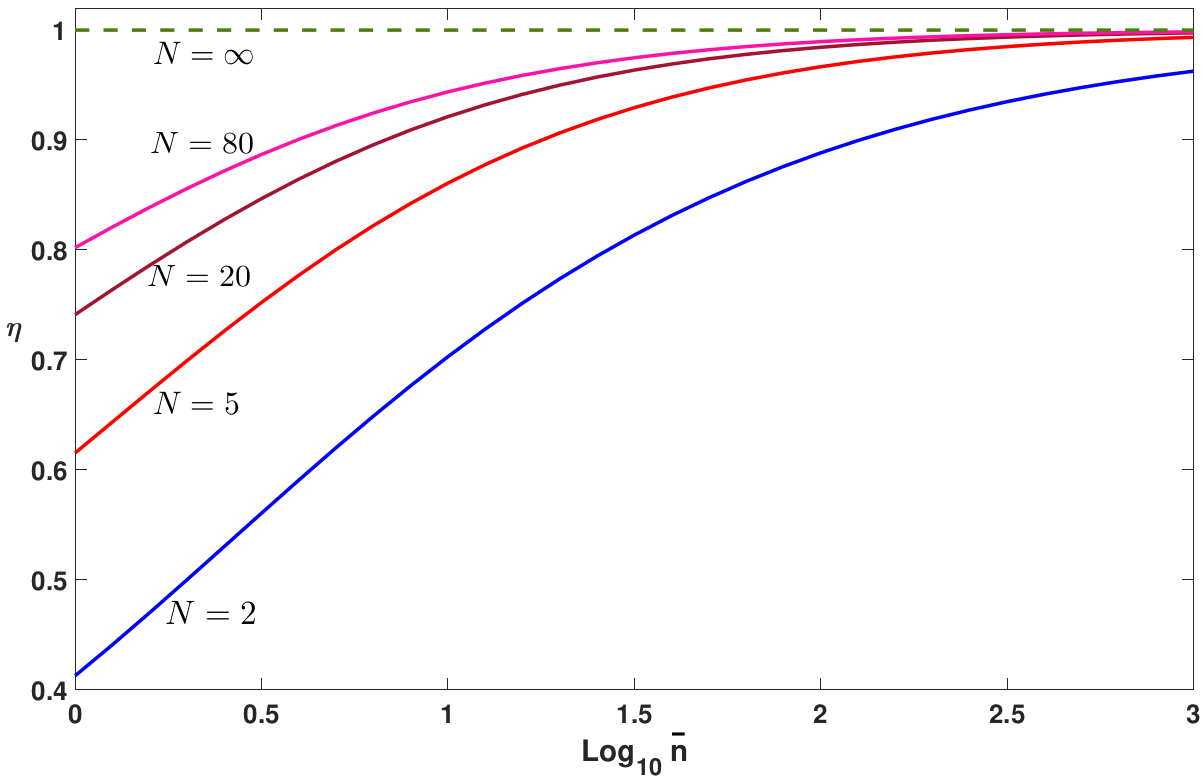}
\caption{\label{f-Carn2}
Scaling of the work extraction efficiency $\eta$ with the number of modes $N$ as a function of the initial mean number of quanta $\bar n$ in a single hot mode and a ``bath'' of $N-1$ initially empty modes. For $N \to \infty$, $\bar n_c \to 0$ limit has been taken.
}
\end{figure}

The scaling of the extractable work as a function of the initial $\bar n$ with $N-1$ empty  modes ($\bar n_c =0$) shows (Fig. \ref{f-Carn2}) that
for  $N\to \infty$ the entire thermal energy can be converted into work, $W\to \hbar \omega\bar n$, as in a Carnot engine with the cold bath at zero temperature.
In the classical limit $\bar n \gg 1$, we find
\begin{eqnarray}
\bar n_f \approx \bar n^{\frac{1}{N}} \bar n_c^{\frac{N-1}{N}}.
\end{eqnarray}
 Namely, the resulting mean number of quanta per mode is the geometric mean of the input mean quanta numbers.
In particular, for $N=2$ one finds
\begin{eqnarray}
\bar n_f \approx \sqrt{\bar n \bar n_c} , \qquad T_f \approx \sqrt{T T_c}.
\end{eqnarray}
This expression coincide with the example analyzed by Landau-Lifschits \cite{landaubook,Greinerbook} for the work available from two finite baths of equal heat capacity and different temperatures.

To realize such a cycle for work extraction, we resort to the combined thermodynamics theorem for a harmonic oscillator in the form (App. \ref{AppendixB})
\begin{eqnarray}
dE = TdS + \hbar \bar n d\omega.
\end{eqnarray}
Namely, work, which corresponds to energy exchange with constant entropy \cite{Kurizkibook}, can be extracted by varying the frequency of the oscillator. 
The process can be accomplished in two steps: 
\begin{enumerate}
\item
{\it Bring all the oscillators to the same temperature adiabatically.} The oscillators keep their mean quanta numbers, but change their frequencies. Their entropies do not change and work is extracted by reducing the frequency of the hot oscillators while spending less work on increasing the frequencies of the cold oscillators.
Let us consider work done in this step for two modes (see Fig. \ref{schemecarnot}) with mean quanta $\bar n$ and $\bar n_c$ for the hot and cold modes, respectively, starting at the same frequency $\omega$. After the adiabatic step, the frequencies are
$\omega_h$ and $\omega_c$, respectively.
The work done on the hot and cold modes in this adiabatic step is given by
\begin{eqnarray}
 W_{h[0\to 1]}&=&\hbar \bar n (\omega_h-\omega)\\ \nonumber
 W_{c[0\to 1]}&=& \hbar \bar n_c (\omega_c-\omega).
\end{eqnarray}
There is no heat exchange between the modes in this step, and they are brought to a temperature $T$ at the end of the step.
\item
{\it Bring all the oscillators to mutual thermal contact and restore their frequencies to the initial value.} This step must also be very slow, so that it occurs reversibly and no entropy is produced. During this step, heat flows from the oscillators whose frequencies increase to the oscillators whose frequencies decrease. Additional net work can be obtained in this step due to the free energy change (see App. \ref{AppendixB}).

For two modes as above, the frequencies are brought back to the initial value $\omega$, and both of the modes finally have mean quanta $\bar n_f$ following Eq. (\ref{Ebalance}), i.e.,
\begin{equation}
 2 S(\bar n_f) = S(\bar n)+S( \bar n_c).
\end{equation}
The work done on the modes is then given by the free energy-difference
\begin{eqnarray}
 W_{h[1 \to 2]}&=&\mathcal{F}_{h,2}-\mathcal{F}_{h,1}= \hbar (\omega \bar n_f- \bar n \omega_h)- T[S(\bar n_f)-S(\bar n)]\\ \nonumber
 W_{c[1 \to 2]}&=& \mathcal{F}_{c,2}-\mathcal{F}_{c,1}=\hbar (\omega \bar n_f- \bar n_c \omega_C)-T[S(\bar n_f)-S(\bar n_C)].
\end{eqnarray}
In this step the hot mode releases heat in the amount
\begin{equation}
 Q_{h[1 \to 2]}= T[S(\bar n)-S(\bar n_f)],
\end{equation}
and the cold mode absorbs the amount of heat
\begin{equation}
 Q_{c[1 \to 2]}= T[S(\bar n_f)-S(\bar n_c)].
\end{equation}

The total work done on the system can be negative, 
\begin{equation}
 W= -\hbar \omega[\bar n + \bar n_c- 2 \bar n_f],
\end{equation}
i.e. net work can be extracted from the system.
\end{enumerate}

The ratio of the extracted work to the hot mode
energy, which can be considered as the efficiency of the work extraction schemes explored in this article, 
\begin{equation}
 \label{rever-effi}
 \eta= \frac{-W}{\hbar \omega \bar n},
\end{equation}
  evaluates for $N$ modes to
\begin{equation}
 \label{rever-effi2}
 \eta= \frac{\bar n + (N-1)\bar n_c- N \bar n_f}{ \bar n}.
\end{equation}
For work extraction one needs $\bar n> \bar n_c$. Additional work can be extracted from the remaining $N$ modes with mean quanta $\bar n_f$ using colder modes $\bar n_c<\bar n_f.$
For $N\to \infty$ we have
\begin{equation}
 \label{rever-effi2}
\eta= 1-\frac{\bar n_c}{\bar n}\left( 1 + \ln \frac{\bar n}{\bar
n_c}\right)
\end{equation}
For $\bar n, \bar n_c \gg1$ or $\bar n\gg 1$ and  $\bar n_c \to 0$, this is equivalent to 
\begin{equation}
\eta_{N \to \infty} \approx 1-\frac{T_c}{T_h}\left( 1 + \ln \frac{T_h}{T_c}\right).
\end{equation}

This section has shown the possibility of efficient  reversible work extraction  from one hot mode and few cold modes by adiabatic manipulations.
Such reversible work extraction, without entropy production, requires infinite time allocation to the two strokes (steps) and thus yields zero power.
Irreversibility, which is needed for finite power generation, tends to lower the efficiency, as is known for conventional heat engines \cite{Curzon_1975,novi}. Another practical difficulty of the present scheme is the need to appreciably change the mode frequencies. These drawbacks prompt us to resort to measurements in Sec. \ref{secIV}.

\section{Work extraction by observation and feedforward (WOF) from  a thermal mode}
\label{secIV}

\noindent To circumvent the tradeoff between efficiency and power which is inherent in heat engines as well as in the reversible scheme of Sec. \ref{secIII}, one may consider information engines (IE) whose power is determined by the measurement duration and the detector resetting time. 

For the minimal scheme of a single hot mode considered here, the need for IE arises since its passive (particularly thermal) state cannot be used for extracting work by unitary transformations (Sec. \ref{secII}). Namely, we need to measure the state and apply feedforward to extract work from the information. We shall present our idea concerning work extraction from a thermal state via quantum measurements that probe only a {\it small fraction} of the input so as to minimize the measurement cost and feedforward of the acquired information in order to steer the unmeasured (dominant) fraction at low cost.

Several methods based on what we have dubbed \cite{WOF} ``work by observation and feedforward (WOF)'' will be compared:

A) WOF by energy measurement of the entire thermal field;

B) WOF by small-fraction photocount; and

C) WOF by small-fraction homodyning.

\subsection{WOF by energy measurement of the entire thermal field}
\label{secIVA}
\noindent If one performs sufficiently many energy measurements of the thermal oscillator mode  and transforms each time the post-measured state to the ground state, the work extracted from the ensemble is the average energy of the oscillator,
\begin{equation}
\langle E \rangle=\hbar \omega \bar n.
\end{equation}
The ideal extraction method consists of many quantum non-demolition (QND) Fock-state measurements, i.e. $|n\rangle \langle n|$ projectors, each projection followed by displacement (downshift) to the ground (vacuum) state $|0\rangle$ via the unitary operation $|0\rangle \langle n|+|n\rangle \langle 0|$. However, such operations are hard to implement.

One must also account for the cost: the heat-up of the detector (assuming it is kept in an environment at temperature $T_D$), by the amount
\begin{equation}
 Q_D=k_BT_D I_D
\end{equation}
that is proportional to its  entropy increase $I_D$. 

In what follows we do not discuss QND operations, but rather measurements with dispersion (spread) in the number state-basis. We then find (c.f. App. \ref{AppendixF}) that the detector entropy increase is the same as the entropy of the input thermal distribution (App. \ref{app:A})
\begin{equation}
  I_D= S(\bar n)/k_B=1+\ln \bar n,~\mbox{for}~\bar n \gg 1.
\end{equation}
The net work gained is therefore
\begin{equation}
 W= \langle E \rangle-Q_D.
\end{equation}


Thus, for $T_D\simeq T_h$, WOF by energy measurement of the entire input is an {\it inefficient method} that wastes most of the gained work on the detector heat up. The only way to gain work by this method is to lower the temperature $T_D$ compared to the input (hot-mode) temperature $T_h$ in order to achieve $W\gg Q_D.$
As discussed in this section, the cost of feedforward required to extract work from the post-measured state increases depending on the information gain.
 To remedy these drawbacks, we proposed \cite{WOF} to lower the entropy increase of the detector (and not only to reduce the temperature $T_D$ to minimize the resetting cost) by measuring only a small fraction of the input, as shown in Sec. \ref{secIVB}, Sec. \ref{secIVC} and Sec. \ref{secVA}.

\subsection{WOF by small-fraction photocount}
\label{secIVB}

\begin{figure}
\includegraphics[width=8cm]{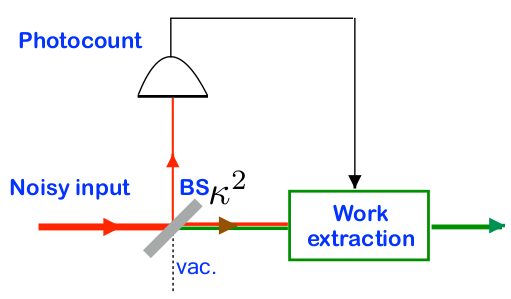}
\caption{\label{photocount}
Scheme of work extraction by quanta number detection (photocount) from a small fraction of a  noisy input signal. The input is incident on a BS with high transmissivity $\kappa^2$, and the photocount is done on the reflected part. Work is extracted by unitary manipulation from the post-measured transmitted part 
}
\end{figure}

\noindent Here, we study work extraction from a thermal field mode by detecting a small fraction of the input in the Fock basis and extracting work from the post-measured state. For an electromagnetic (EM) field mode, this corresponds to photocounts performed on the sampled (reflected) fraction (Fig. \ref{photocount}) of a thermal input incident on a beam splitter (BS) with high transmissivity $\kappa^2$.

\begin{figure}
\begin{center}
\includegraphics[width=0.45\linewidth]{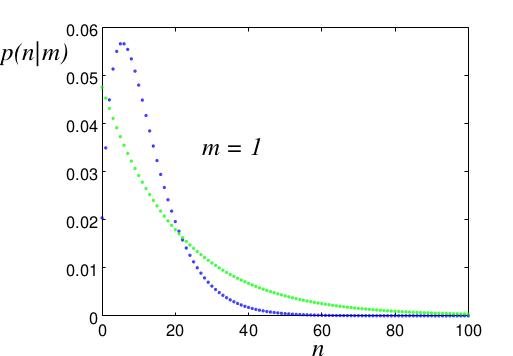}
\includegraphics[width=0.45\linewidth]{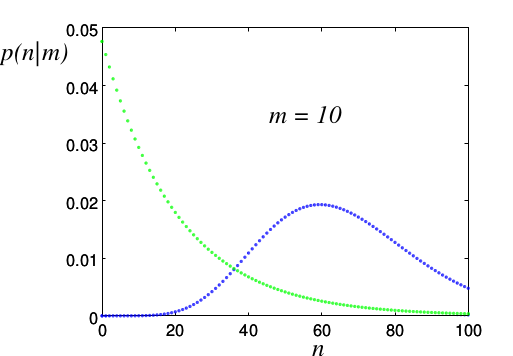}
 \caption{ \label{plotpm} The probabilities $p(n|m)$ (blue dots) of occupying the $n$th Fock state in the post-measured (Eq. \ref{fockpm}) state when the measurement outcome is $m$ quanta for a thermal input with mean number of quanta $\bar n=20$, and a BS with transmissivity $\kappa^2=0.9$. The plots clearly show that the occupation probabilities are non-monotonic and therefore the post-measured state is non-passive. The green dots represent the quanta number distribution of the thermal input with mean number of quanta $\bar n=20$.
}
\end{center}
\end{figure}

\begin{figure}
\begin{center}
\includegraphics[width=0.95\linewidth]{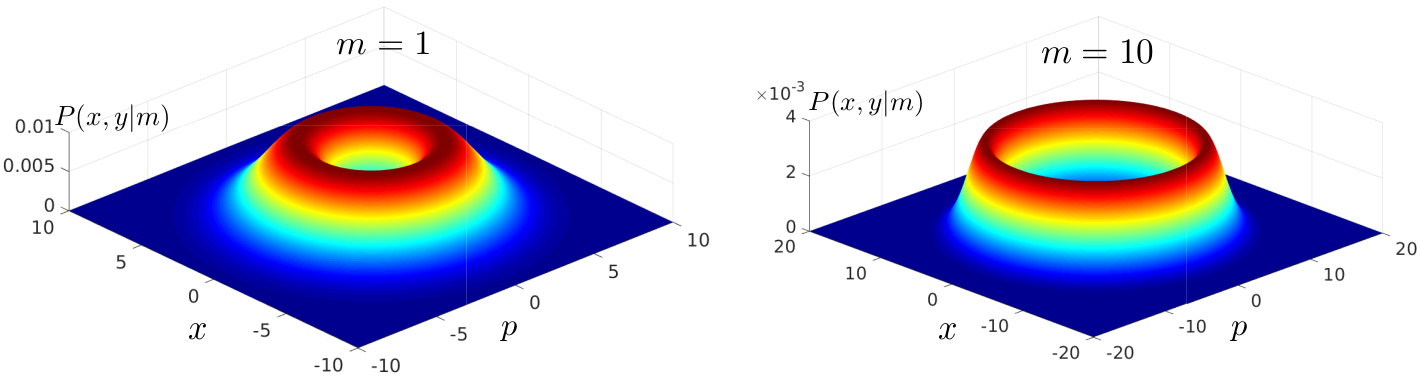}
 \caption{ \label{plotpdis} The P-distribution of the post-measured state for a thermal input with $\bar n=20$ when the measurement outcome is $m$ quanta (App.\ref{app:D}): $m=1$ (left) and  $m=10$ (right). The BS transmissivity $\kappa^2=0.9$. The plots clearly show the non-monotonicity of the distribution which implies non-passivity. 
}
\end{center}
\end{figure}

A thermal state at the input temperature $T_h$ (with mean quanta number $\bar n$) in the Fock basis is represented as
\begin{equation}
\label{fockthermal}
 \rho(T_h)= \sum_n p_n |n\rangle \langle n|,
\end{equation}
where $p_n=e^{-\frac{\hbar \omega n}{k_B T}}(1-e^{-\frac{\hbar \omega }{k_B T}})$ is the occupancy of the $n$th Fock state. 
Detection of $m-$quanta of the reflected beam occurs with a probability (App. \ref{app:C})
\begin{equation}
 p_m= (1-e^{-\frac{ \hbar \omega}{k_B T'}})e^{-\frac{ \hbar \omega}{k_B T'} m},
\end{equation}
where we have introduced the effective temperature of the reflected beam
\begin{equation}
\label{effective-T}
 \exp \left( \frac{\hbar \omega}{k_BT'}\right)= \frac{\exp \left( \frac{\hbar \omega}{k_BT}\right)-\kappa^2}{1-\kappa^2}.
\end{equation}
This $m-$quanta detection yields the post-measured (conditional) state that has transmitted through the high transmissivity BS
\begin{equation}
\label{fockpm}
 \rho_{m}= \frac{1}{m!}(1-e^{-\frac{\hbar \omega }{k_B T}}\kappa^2)^{m+1}\left[\sum_{n=0}^\infty e^{-\frac{\hbar \omega }{k_B T} n}(\kappa^2)^{n} \frac{(n+m)!}{n!}|n\rangle \langle n|\right]=\sum_{n=0}^\infty p(n|m)|n\rangle \langle n|.
\end{equation}
The thermal distribution is modified in the post-measured state depending on $\bar n$, $\kappa^2$ and the measurement outcome $m$ (App. \ref{app:C}, Fig. \ref{plotpm}). 
The projection on the $m-$ quanta state of the reflected beam can be geometrically represented by a ring in the phase plane which cuts out a hollow crater from the input thermal (Gaussian) distribution
(Fig. \ref{plotpdis}).  We observe the non-monotonicity of the post-measured state distribution which attests to non-passivity in Fig. \ref{plotpm}, \ref{plotpdis}, as the state, diagonal in Fock basis, in Eq. (\ref{fockpm}) is passive iff the probabilities satisfy \cite{Allahverdyan_2004_a}
\begin{equation}
\label{passive-condi}
 p(n|m)\geq p(n'|m),~~~\forall n,n',~~~\mbox{when}~~E_n>E_{n'}.
\end{equation}

  The departure from the thermal character of the input is also manifested by the second-order coherence function \cite{Scully_BOOK} for the post-measured state in Eq. (\ref{fockpm}) which can be evaluated to be
\begin{eqnarray}
 g^{(2)}(0)&=&\frac{\langle a^{\dagger^2}a^2\rangle}{\langle a^{\dagger}a\rangle^2} \\ \nonumber
 &=& \frac{\sum_{n=0}^\infty p(n|m)~n (n-1)}{\bar n_m^2}\\ \nonumber
 &=& 1+ \frac{1}{1+m}.
\end{eqnarray}
Thus, the second-order coherence function is {\it independent of the input beam temperature} and the splitting ratio of the BS. When no photon is detected, i.e., the entire beam passes through the BS, the transmitted beam is thermal as expected, with $g^{(2)}(0)=2$. In the limit of a large  number of detected quanta $m\gg 1$, $g^{(2)}(0)$ of the transmitted beam converges to $g^{(2)}(0)=1$, indicating Poissonian statistics.
  
Work cannot be extracted by displacing the post-measured state in Eq. (\ref{fockpm}), as its mean quadratures are zero (Sec. \ref{secII}). Instead, one can perform a (unitary) permutation in the Fock-basis such that the modified probabilities satisfy the passive-state condition Eq. (\ref{passive-condi}). The average energy of the post-measured state 
 \begin{equation}
  E_m= \hbar \omega \frac{(1+m)\kappa^2}{e^{\frac{\hbar \omega }{k_B T}} -\kappa^2}
 \end{equation}
 can always be lowered  to $E'_m$ by the permutation that renders the state passive thereby extracting the amount of work
 \begin{equation}
  W_m=E_m-E'_m.
 \end{equation}
Upon averaging over all outcomes $m$, the net work gain can be obtained as
 \begin{equation}
  W=\sum_m p_m W_m.
 \end{equation}

The average information stored in the detector that needs to be erased after WOF is
\begin{equation}
\label{IRPC}
 I_D=-\sum_m p_m \ln (p_m ).
\end{equation}
We show in Sec. \ref{secIV-1A} that the detector heat-up cost $Q_D= k_BT_DI_D$ can be made lower than its counterpart for entire energy measurement in Sec. \ref{secIVA}. The practical merit of small-fraction photocount is that it is much easier to implement than the perfect QND measurement required in Sec. \ref{secIVA}.

The mutual information can be expressed as (App. \ref{app:F})
\begin{eqnarray}
\label{IPC}
I =- \sum_{n} p(n) S(p(m|n)) +S(p(m))
\end{eqnarray}
This average mutual information, which is always non-negative, tells us by how much the energy uncertainty is reduced on average depending on the measurement ($m-$quanta) outcome.


We find that WOF by photocount of a small-fraction can have high efficiency for a suitable BS transmissivity $\kappa^2$. However, the inability to fully sample the phase-space distribution of the input in the Fock basis limits the efficiency. Practically, simple operations such as displacement cannot extract work in this scheme, as noted above. We therefore resort to WOF homodyning.

\subsection{WOF via phase-sensitive (homodyne) measurement of a small fraction}
\label{secIVC}

\noindent In the quantum domain, joint position and momentum measurements cannot be done perfectly and are limited by the quantum uncertainty.
Nevertheless, we have shown that \cite{WOF} a passive (thermal) signal can be used to efficiently extract work via homodyne measurements of its non-commuting orthogonal quadratures performed on a smal-fraction of the input and followed by information feedforward of the unmeasured dominant fraction. The open issue we address is: what is the rapport between work output and information gain in this scheme?

We first briefly present this scheme where the hot field mode is incident on a BS with high transmissivity $\kappa$, and a homodyne measurement is performed on the orthogonal quadratures of a split fraction of the incoming field. The remaining part of the field is projected onto a state from which work can be extracted by a unitary transformation (displacement). To find the maximum extractable work, one has to take into account the energy cost of the measurement and the quantum noise entering the scheme.

\begin{figure}
\includegraphics[width=10cm]{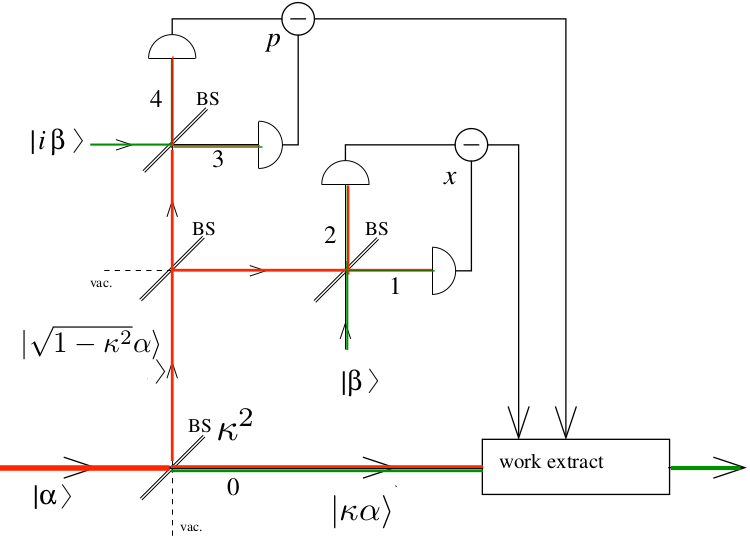}
\caption{\label{schemewof}
Scheme of the setup for WOF by small-fraction homodyning: an unknown state (here labeled as coherent state $|\alpha \rangle$) enters a beam splitter which transmits a fraction $\kappa^2$ of the input energy and reflects $\sqrt{1-\kappa^2}$. On the reflected part, a homodyne measurement is performed to estimate the quadratures $x$ and $p$.
}
\end{figure}

A thermal state of harmonic oscillator can be represented as a mixture of coherent states $|\alpha\rangle$. Assume first that a coherent state $|\alpha\rangle$ with complex coherent amplitude $\alpha=\frac{1}{\sqrt{2}}(x+ip)$ enters the setup in Fig. \ref{schemewof}. After the first BS with splitting ratio $\kappa^2/(1-\kappa^2)$, the state $|\kappa \alpha\rangle$ is transmitted and the state $|\sqrt{1-\kappa^2} \alpha\rangle$ is reflected towards the homodyne detectors for estimating the quadratures $\hat x$
and $\hat p$ of the input state.
We resort to a local oscillator in a coherent state with real quadrature-amplitude $\beta$ and to its imaginary quadrature counterpart with amplitude $i\beta$. The modes $0,1,2,3,4$ behind the BS are in a multimode (product) coherent state.
The photocount differences $\Delta n_x \equiv n_1-n_2$ and  $\Delta n_p \equiv n_{3}-n_{4}$ in Fig. (\ref{schemewof}) carry information on the input-field quadratures $x$ and $p$ \cite{WOF}.

 Let us now take the input state to be a mixture of coherent states,
\begin{eqnarray}
\hat \varrho = \int \!  \int P(\alpha) |\alpha \rangle \langle \alpha | d^2 \alpha .
\end{eqnarray}
For a thermal input state with mean number of quanta $\bar n$, the Glauber-Sudarshan phase-space distribution is Gaussian in the quadratures
\begin{eqnarray}
P(\alpha) = \frac{1}{\pi \bar n}\exp \left(- \frac{|\alpha|^2}{\bar n}  \right)\equiv P(x,p) &=& \frac{1}{2\pi \bar n}\exp \left(- \frac{x^2+p^2}{2\bar n}  \right).
\end{eqnarray}
The distribution of $\alpha$, conditioned on the detection of quanta number differences  $\Delta n_{x}$ and $\Delta n_{p}$, is \cite{WOF}
\begin{eqnarray}
 P(\alpha |\Delta n_{x},\Delta n_{p}) = \frac{p(\Delta n_{x},\Delta n_{p}|\alpha) P(\alpha) }{p(\Delta n_{x},\Delta n_{p})}.
\label{PalcondDn}
\end{eqnarray}
\begin{figure}
\begin{center}
\includegraphics[width=0.95\linewidth]{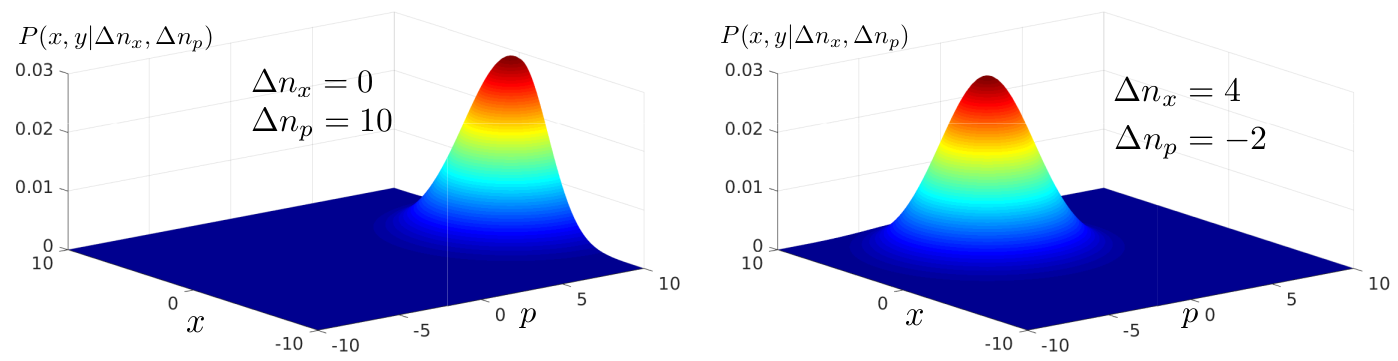}
 \caption{ \label{plotpdis-homo} The P-distribution of the post-measured state for a thermal input with $\bar n=16$ when the measurement outcomes are $\Delta n_{x}$ and $\Delta n_{p}$. The BS transmissivity $\kappa^2=0.75$ and $\beta=1.4$.
}
\end{center}
\end{figure}
The unmeasured (transmitted) field mode has the state (conditional on the detection of  $\Delta n_{x}$ and $\Delta n_{p}$)
\begin{eqnarray}
\hat \varrho (\Delta n_{x},\Delta n_{p}) 
&=& \frac{1}{\kappa^2} \int \! \int P\left(\frac{\alpha}{\kappa}|\Delta n_{x},\Delta n_{p}\right) |\alpha \rangle \langle \alpha | d^2 \alpha.
\label{varrhoDn}
\end{eqnarray}
This state has in general {\it non-vanishing mean values} of quadratures $\hat x$ and $\hat p$,
\begin{eqnarray}
\label{eqmeanX}
\langle \hat x\rangle &=&\kappa \int \! \int x P(\alpha|\Delta n_{x},\Delta n_{p})  d^2 \alpha , \\ \nonumber
\langle \hat p\rangle 
&=&\kappa \int \! \int p P(\alpha|\Delta n_{x},\Delta n_{p})  d^2 \alpha .
\label{eqmeanP}
\end{eqnarray}
A great merit of this scheme is that
one can extract most of the stored work (albeit not fully \cite{WOF}) by simply downshifting (displacing to the origin) the state ( \ref{varrhoDn}) such that the mean quadratures of the final state are zero. 
The mean work obtained in this process can be found by averaging $\frac{\hbar \omega}{2}\left(\langle \hat x\rangle ^2 + \langle \hat p\rangle ^2 \right)$ over all values of $\Delta n_{x},~~\Delta n_{p}$ and subtracting the invested energy of the two local oscillators $2 \hbar \omega\beta^2$. 
The extractable work is then found to be
\begin{eqnarray}
\label{wdis}
W &\approx& \frac{\hbar \omega}{2}\int \int \left(\langle \hat x\rangle ^2 + \langle \hat p\rangle ^2 \right)
p(\Delta n_{x},\Delta n_{p}) d\Delta n_{x}d\Delta n_{p} -2 \hbar \omega\beta^2 \\ \nonumber
&\approx& 2 \hbar \omega\beta^2\left[\frac{\kappa^2 (1-\kappa^2) \bar n^2}{2\beta^2 +(1-\kappa^2)(1+2\beta^2)\bar n} -1\right].
\end{eqnarray}
The expression can be optimized with respect to $\beta$ and $\kappa$ (App. \ref{app:E}).
The resulting maximal work gained by using the information as feedforward to downshift the unmeasured part is
\begin{eqnarray}
W_{\rm max} \approx
\hbar \omega\left(\sqrt{\bar n - \sqrt{\bar n}+1}-1 \right)^2  \left( 1-\frac{1}{\sqrt{\bar n}} \right).
\label{Woptimum}
\end{eqnarray}
\begin{figure}
\includegraphics[width=10cm]{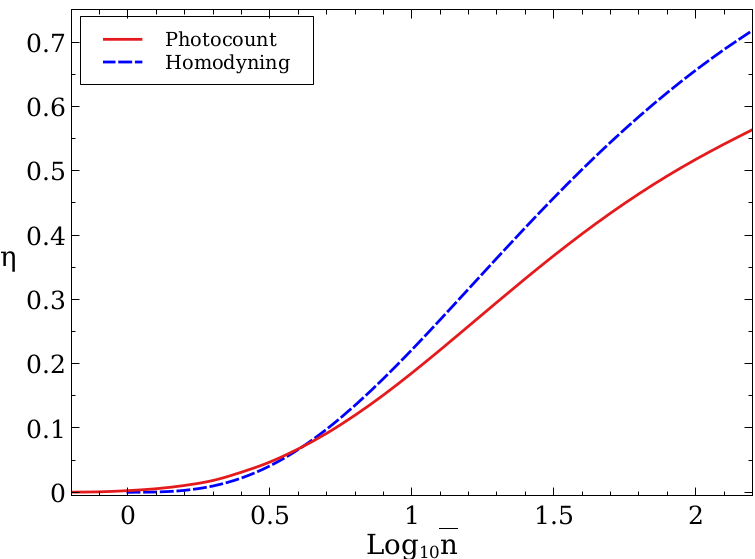}
\caption{\label{photowork}
Efficiency of extractable work by small-fraction photocount WOF compared to small-fraction homodyne WOF as a function of the mean input quanta number $\bar n$. The BS transmissivity for the photocount scheme is $\kappa^2=0.75$.
}
\end{figure}

\begin{figure}
\includegraphics[width=10cm]{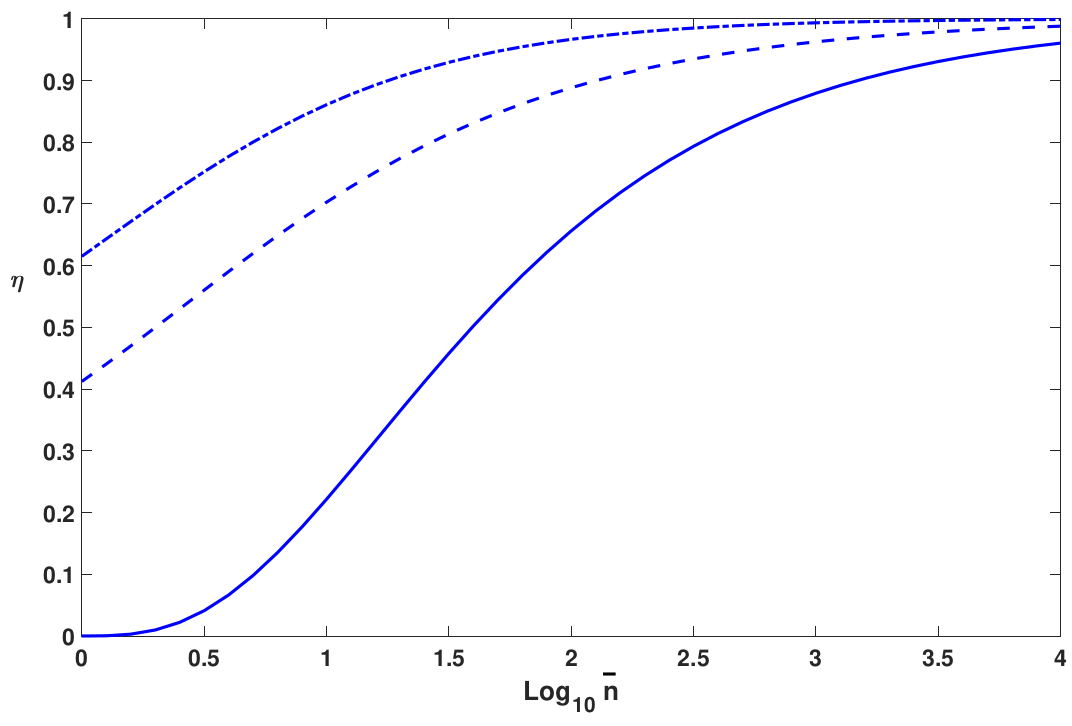}
\caption{\label{f-Carn5}
Comparison of the efficiency (solid) in the WOF homodyning scheme with the reversible scheme of Sec. \ref{secIII} for $N=2$ (dashed) and $N=5$ (dashed-dotted) as a function of mean input quanta $\bar n$. All but one modes are assumed to be initially empty.
}
\end{figure}
For large $\bar n$ the optimized values of the BS transmissivity and local oscillator energy read as $1-\kappa^2= \frac{1}{\sqrt{\bar n}}$ and  $2 \beta^2= \sqrt{\bar n}$, respectively, and the maximal extractable work is given by
\begin{eqnarray}
W_{\rm max} \approx
\hbar \omega\left[ \bar n- 4\sqrt{\bar n}+ 6+ O(\frac{1}{\sqrt{\bar n}},\frac{1}{\bar n})\right].
\label{Woptimum-large-n}
\end{eqnarray}

In Fig. \ref{photowork} we compare the efficiency in the small-fraction WOF homodyning scheme with that of small-fraction photocount in Sec. \ref{secIVB}. In photocount do not obtain phase  (coherence) information, only number-state probabilities, so that the work extraction and its efficiency are expected to be lower than by homodyning. However, this turns out to be true only for large $\bar n$. This is due to the local oscillator energy invested in homodyning WOF, whereby much of the work is wasted for small $\bar n$,  in contrast to photocount.

In Fig. \ref{f-Carn5} we perform a similar comparison with the reversible scheme of Sec. \ref{secIII}. It is seen that the two schemes are comparable for $\bar n \gg 1$.

Let us calculate the information gain, characterized by the mutual information \cite{Vid2016,SagawaPRL08}, that needs to be processed for feedforward in this small-fraction homodyne WOF. The average mutual information can be expressed as (App. \ref{app:F})
\begin{eqnarray}
I&=&\int \int \int \int \ln \frac{P(x,p |\Delta n_x, \Delta n_p)}
{P(x,p)} P(x,p |\Delta n_x, \Delta n_p) p(\Delta n_x, \Delta n_p)d\Delta n_x d\Delta n_p dx dp\\ \nonumber
&=& \langle I_x \rangle + \langle I_p \rangle,
\label{EqP1}
\end{eqnarray}
as the joint probabilities factorize 
 in $x$ and $p$, where
\begin{eqnarray}
\langle I_x \rangle =  \int \int \ln \frac{P(x |\Delta n_x)}
{P(x)} P(x, |\Delta n_x)p(\Delta n_x) d\Delta n_x  dx =\langle I_p \rangle  
\end{eqnarray}
Using the optimized values of $\beta$ and $\kappa$ for maximum work extraction (App. \ref{app:E}), we obtain
for $\bar n \gg 1$ 
\begin{eqnarray}
 \langle I_x \rangle  &\approx&
-\frac{1}{2}\ln \left(1-\frac{\bar n}{\bar n + 2\sqrt{\bar n}} \right).
\end{eqnarray}
The total mean mutual information in this approximation is thus
\begin{eqnarray}
\label{IH}
I= \langle I_x \rangle +\langle I_p \rangle\approx \frac{1}{2} \ln \frac{\bar n}{4}. 
\end{eqnarray}

The corresponding cost of signal processing (feedforward) has the lower bound \cite{ffepl}
\begin{equation}
 E_F\geq k_B T_D \frac{1}{2} \ln \frac{\bar n}{4}.
\end{equation}
Therefore, the bound on the cost of feedforward is much lower for large $\bar n$ compared to the work gain which scales with $\bar n$. One can further lower the cost by reducing the environment temperature $T_D$.

\section{Cost of information erasure}
\label{secIV-1}

\subsection{Resetting cost following photocount of entire thermal input}
\label{secIV-1A}
The increase in detector entropy discussed in Sec. \ref{secIVA}
sets a bound on  WOF efficiency by photocount of the entire thermal input
\begin{equation}
 \eta \le \frac{\hbar \omega \bar n- k_BT_D I_D}{\hbar \omega \bar n}.
\end{equation}
For $\bar n\gg 1$, this bound becomes
\begin{equation}
 \eta \le 1- \frac{k_BT_D (1+\ln (\bar n))}{\hbar \omega \bar n}.
\end{equation}
The condition for non-zero WOF efficiency in this scheme is thus
\begin{equation}
\label{lim1}
 k_BT_D\leq \frac{\hbar \omega \bar n}{1+\ln (\bar n)}.
\end{equation}
This bound on $T_D$ will be compared in what follows to its counterpart by small-fraction WOF.
The above bound for non-zero efficiency can also be expressed as (using App. \ref{app:A})
\begin{equation}
 \bar n \geq  \bar n_D + \frac{1}{1-\frac{\ln \bar n_D}{\ln (\bar n_D +
1)}},
\end{equation}
where $\bar n_D$ is the mean number of quanta in the detector when it is kept in an environment at a temperature $T_D$.

\subsection{Resetting cost for small-fraction photocount} 

For the small-fraction photocount scheme in Sec. \ref{secIVB}, the tradeoff between the detection cost (energy and entropy) and the amount of extracted work obviously depends on the reflected fraction $1-\kappa^2$. Inspired by the optimization for homodyne WOF (App. \ref{app:E}), we choose the reflected $1-\kappa^2$ fraction to be 
\begin{equation}
 1-\kappa^2=1/\sqrt{\bar n}.
\end{equation}
In this case the detector uses (absorbs, in the case of photons) $\sqrt{\bar n}$ mean quanta for detection, instead of $\bar n$ without BS.
The mean energy used for detection is $1/\sqrt{\bar n}$ fraction of the input.
The entropy increase of the detector  is given for $\bar n\gg 1$ by 
\begin{equation}
I_D= 1+\ln (\sqrt{\bar n}).
\end{equation}
Therefore, the detector heat-up cost is then
\begin{equation}
 Q_D=k_BT_D(1+ \ln (\sqrt{\bar n})),
\end{equation}
for large $\bar n$.
The corresponding upper bound of WOF efficiency is
\begin{equation}
 \eta \le  \frac{(1-1/\sqrt{\bar n})\hbar \omega \bar n-k_BT_D (1+ \ln (\sqrt{\bar n}))}{\hbar \omega \bar n},
\end{equation}
Hence, for $\bar n \gg 1$ the condition of non-zero WOF efficiency is modified to
\begin{equation}
 \label{lim2}
  k_BT_D\lesssim \frac{2\hbar \omega \bar n}{2+\ln (\bar n)}.
\end{equation}
Therefore,  by resorting to the small fraction photocount in Sec. \ref{secIVB}, we can almost {\it double the upper limit} on $T_D$ for work extraction, which is a considerable advantage.



\subsection{Resetting following small-fraction homodyne WOF}

%
%
In this scheme, the entropy increase of the detectors factorizes for $x$ and $p$, yielding
\begin{equation}
 I_{Dx}=I_{Dp}=-\int P(\Delta n_{x}) \ln P(\Delta n_{x})= \frac{1}{2}[1+\ln (2\pi \sigma_{\Delta n}^2)],
\end{equation}
where
\begin{equation}
  P(\Delta n_{x},\Delta n_{p})\approx 
\frac{1}{2\pi \sigma_{\Delta n}^2} \exp \left[ -\frac{\Delta n_{x}^2+\Delta n_{p}^2}{2\sigma_{\Delta n}^2}  \right]
 ;~~2\sigma_{\Delta n}^2 = 2\beta^2 + \bar n (1-\kappa^2)\left(2 \beta^2 + 1 \right)\approx \bar n+ 2\sqrt{\bar n}
\end{equation}

The total entropy increase is then
\begin{eqnarray}
\label{IRH}
 I_D=2I_{Dx} &=&1+\ln (2\pi \sigma_{\Delta n}^2)\\
 &=&1+ \ln \pi(\bar n+ 2\sqrt{\bar n}).
 \end{eqnarray}
 Therefore, by following the same procedure as for the photocount scheme, we find the upper bound on $T_D$ for non-zero WOF efficiency to be
 \begin{equation}
 k_BT_D<\frac{\hbar \omega \bar n}{1+\ln (\pi \bar n)},
 \end{equation}
 for $\bar n \gg 1$.

If, instead of small-fraction homodyne, one performs homodyning on the entire field, one has
\begin{eqnarray}
 P'(\Delta n_{x},\Delta n_{p})\approx 
\frac{1}{2\pi \sigma^2} \exp \left[ -\frac{\Delta n_{x}^2+\Delta n_{p}^2}{2\sigma^2}  \right],
\label{EqPn2}
\end{eqnarray}
where 
\begin{equation}
2\sigma^2=2\beta^2 + \bar n \left(2 \beta^2 + 1 \right)\approx \bar n \sqrt{\bar n}+ \bar n+\sqrt{\bar n} 
\end{equation}
for large $\bar n$.
The entropy increase of the detector in this case
\begin{eqnarray}
 I_D= 1+\ln \pi(\bar n\sqrt{\bar n}+\bar n+ \sqrt{\bar n}),
 \end{eqnarray}
assuming the same local oscillator energy as for small-fraction homodyning. The upper bound on $T_D$ for non-zero WOF efficiency is then given by
 \begin{equation}
 k_BT_D<\frac{\hbar \omega \bar n}{1+\ln (\pi \bar n^{3/2})},
 \end{equation}
 for $\bar n \gg 1$. This implies that one can increase the upper bound on $T_D$ almost by a factor of $3/2$ by resorting to small-fraction homodyne WOF, instead of the entire field homodyne scheme.

\subsection{Entropy increase of the detectors}

\noindent We can estimate $\Delta \bar n_D$, the increase in the mean quanta number from its initial mean number $\bar n_D$ for a detector kept in an environment at temperature $T_D$, to be
\begin{eqnarray}
\label{deten}
\Delta \bar n_D=  \frac{1-\kappa^2}{4}\bar n + \frac{1}{2}\beta^2.
\end{eqnarray}

The total entropy increase in the $4$ detectors of homodyne WOF for the thermal (highest-entropy) state with the mean quanta number $\bar n_D+\Delta \bar n_D$ is, in bits
\begin{eqnarray}
\Delta S_D \approx 4 S(\Delta \bar n_D+\bar n_D)-S(\bar n_D),
\end{eqnarray}
where 
\begin{equation}
S(\bar n_D)=k_B[(\bar n_D +1)\ln (\bar n_D +1) - \bar n_D \ln \bar n_D]
\end{equation} 
is the entropy of each detector corresponding to $\bar n_D$

In the large-$\bar n$ limit one then has, per detector
\begin{eqnarray}
\label{79}
 \Delta \bar n_D&=&\frac{\sqrt{\bar n}}{2},\\ \nonumber
 \Delta S_D &\approx & S(\Delta \bar n_D)\approx k_B[1 + \ln \Delta \bar n]\\ \nonumber
 \end{eqnarray}
 Equation (\ref{79}) shows that, since only a fraction $\sim 1/\sqrt{\bar n}$ of the input is detected by WOF,
the total entropy change in the four detectors is then
\begin{equation}
\Delta S_D \approx \frac{k_B}{2} \ln \frac{\bar n}{4}.
\end{equation}

\subsection{Energetically optimal resetting cost}

\noindent Let us consider 
an energetically optimal strategy to reset the photodetectors modeled as oscillators. To this end, we assume that we can control the detector frequency (energy gap) (see Fig. \ref{f-zeromemory}) and implement the following steps:
\begin{figure}
\includegraphics[width=10cm]{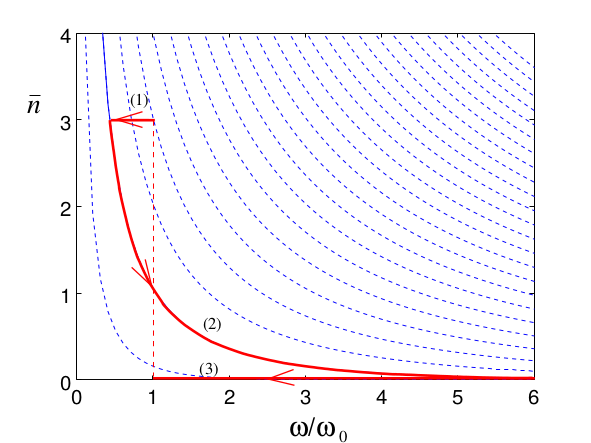}
\caption{\label{f-zeromemory}
Resetting the detectors with energy gap $\hbar \omega$ (in units of $\hbar \omega_0$):The evolution is marked by red line in the $(\omega,\bar n_D)$ plane. (1) Starting with $\bar n_D=3$, adiabatically decrease $\omega$ to reach the initial temperature. (2) Isothermally increase $\omega$ to reach $\bar n_D=0$. (3) Adiabatically return the frequency to its initial value. Broken blue lines are isotherms at various temperatures.
}
\end{figure}
\begin{enumerate}
\item
Adiabatically decrease the frequency $\omega'$ of the detector mode until the detector equilibrates  with the environment at temperature $T_D$. This requires
\begin{eqnarray}
\omega' = \frac{k_BT_D}{\hbar}\ln \left( 1+ \frac{1}{\bar n_D} \right)        .
\end{eqnarray}
During this step one can get work in the following amount from the 4 detectors in homodyne WOF
\begin{eqnarray}
W_1 &=& 4\hbar(\omega - \omega')\bar n_D \nonumber \\
&=& 4\hbar \omega \bar n_D - 4kT_D \bar n_D \left[\ln (\bar n_D+1) - \ln \bar n_D \right].
\end{eqnarray}
\item
Isothermally increase their frequency  to such a value that the mean photon number in the mode vanishes, i.e., $\hbar \omega_f \gg k_BT_D$. 
To this end one has to perform the work 
\begin{eqnarray}
W_2 &=& 4\hbar \omega' \frac{\ln (1+ \bar n_D)}{\ln \left(1+\frac{1}{\bar n_D} \right)} \nonumber \\
&=& 4k_BT_D \ln (1+ \bar n_D).
\end{eqnarray}
The heat dissipated to the environment by the 4 detectors is then
\begin{eqnarray}
Q_D &=& 4\hbar \omega' \left[\bar n_D + \frac{\ln \left( 1+\bar n_D \right)}{\ln \left( 1+\frac{1}{\bar n_D}\right)} \right] \nonumber \\
&=& 4k_BT_D \left[\bar n_D \ln \left( 1+\frac{1}{\bar n_D}\right)+  \ln \left( 1+\bar n_D \right) \right]  \nonumber \\
&=& 4k_BT_D \left[(\bar n_D +1)\ln (\bar n_D +1) - \bar n_D \ln \bar n_D
\right].
\label{EqQLandauer}
\end{eqnarray}
\item
Adiabatically bring the frequency of the oscillator to its initial value. Since no quanta are present at this stage, this action requires no work.
\end{enumerate}
Thus, the work required for resetting the detectors is
\begin{eqnarray}
W_{\rm R} &=& W_2-W_1 \nonumber \\
&=& 4k_BT_D \left[(\bar n_D +1)\ln (\bar n_D +1) - \bar n_D \ln \bar n_D
\right] - 4\hbar \omega \bar n_D.
\label{EqWreset}
\end{eqnarray}
From Eqs. (\ref{EqQLandauer}), (\ref{EqWreset}), we have
\begin{eqnarray}
\label{700}
Q_D =  4\hbar \omega \bar n_D + W_{R}.
\end{eqnarray}
Eq. (\ref{700}) shows that the heat dissipated by the detector resetting is partly covered by the energy stored in the detectors, $ 4\hbar \omega \bar n_D$, and partly by additional work, $W_{R}$, that needs to be invested in the resetting.
This additional work can however be zero for $\bar n_D$ satisfying
\begin{eqnarray}
\frac{\hbar \omega}{k_BT_D} = \left( 1+ \frac{1}{\bar n_D} \right)\ln (\bar n_D+1) - \ln \bar n_D.
\end{eqnarray}
For $\bar n_D$ higher than this value the net work $W_{\rm R}$ is negative. Namely, one can get useful work by resetting the detectors to zero by manipulating the detector frequency.

While the outlined method may, in principle, save us energy or work on the detector resetting, it suffers from the same drawbacks as the work extraction scheme in Sec. \ref{secII}. It is adiabatic, i.e. extremely slow, and requires frequency manipulation of the detectors modeled as oscillators. Yet, it is preferable to reset the detectors by continuously cooling them at the highest rate possible, since the work consumption is modest provided the initial detector temperature is low enough. It is particularly important to maximize the WOF power, which is limited by the detector cooling time. State-of-the-art superconducting photodetection allows ns-scale detector resetting by cooling \cite{Natarajan,Wolff2020}.

\section{Work extraction from partial information: Coarse graining effects}
\label{secV}

\subsection{Why consider coarse graining?}
\noindent For practical reasons, detectors may not have sufficient resolution to record the full information available on the input state, either by photocounts or homodyning. This situation prompts a conceptual question: how does the tradeoff between resolution and information affect the extractable work efficiency?

The distribution of photocounts in each detector, for large quanta numbers, can be well approximated by the Gaussian distribution. The question is: how does this distribution change under coarse graining?
The Gaussian distribution of a random variable $x$ has the form $G(x)= \frac{1}{\sigma \sqrt{2 \pi}} e^{-\frac{1}{2}(\frac{x-\mu}{\sigma})^2}$.
We take, as is customary,  the continuous limit of the photocount probability function (although the counts are discrete). We assume the coarse grained detector to be such that it cannot differentiate between counts of photocounts in blocks of size $R$. We set the blocks such that the mean of the distribution is in the middle of a block. As an example, The probability that an outcome is in a block which is, say, $M$ blocks to the right from the mean is given by
\begin{equation}
 \int_{r1}^{r2} G(x) dx= \pi ({\rm Erf}[\frac{\mu-r_1}{\sqrt{2}\sigma}]-{\rm Erf}[\frac{\mu-r_2}{\sqrt{2}\sigma}]),
\end{equation}
where $r_1= \mu +(M+1/2)R$, $r_2= \mu +(M+3/2)R$ and the error function $
 {\rm Erf} (x)=\int_0^x e^{-t^2} dt$.
The protocol is then as follows:
\begin{itemize}
 \item Assume the resolution of the detectors is $R$, for the $\Delta n_{x/p}$ records assume the values $0, R, 2R,...NR$.
 \item Calculate the average post-measured state and the corresponding work extractable by displacement from each block of area $R\times R$ to get the average work gain from the coarse grained WOF.
\end{itemize}
%

\subsection{Extremely coarse grained homodyning: WOF via sign measurements}
\label{secVA}
\noindent Let us consider an extreme coarse-grained situation where the detected signal ( $\Delta n_x$ or $\Delta n_p$) 
is positive ($+$) or negative ($-$). 
There are four distinct possibilities for sign( $\Delta n_x$) and sign( $\Delta n_p$), corresponding to work gain by displacement $W_{++}$, $W_{+-}$, $W_{-+}$, $W_{--}$: For e.g., $W_{+-}$  we get
\begin{equation}
W_{+-}=\frac{\hbar \omega}{2}(\langle \hat x\rangle_{+-}^2+ \langle \hat p\rangle_{+-}^2)=\hbar \omega\frac{\kappa^2}{16\gamma^2}\frac{2 \sigma^2_{\Delta n}}{\pi}.
\end{equation}
Here we have used
\begin{eqnarray}
 \langle \hat x\rangle_{+-}&=& \kappa \int_{-\infty}^{-\infty}dx \int_{-\infty}^{-\infty}dp\int_{-\infty}^{0}d\Delta n_{p} \int_{0}^{\infty}d\Delta n_{x}~x P(x,p |\Delta n_{x},\Delta n_{p})P(\Delta n_{x},\Delta n_{p}) \\ \nonumber
 &=& \frac{\kappa}{4\gamma}\sqrt{\frac{2 \sigma^2_{\Delta n}}{\pi}}, \\ \nonumber
  \langle \hat p\rangle_{+-}&=&- \frac{\kappa}{4\gamma}\sqrt{\frac{2 \sigma^2_{\Delta n}}{\pi}};~\gamma=\beta \sqrt{1-\kappa^2}\left[1+\frac{1}{\bar n (1-\kappa^2)}+\frac{1}{2\beta^2} \right];~\sigma_{\Delta n}^2 = \beta^2 + \bar n (1-\kappa^2)\left( \beta^2 + \frac{1}{2} \right). 
\end{eqnarray}
%
The total average work obtained by downshifting the post-measured state following a sign measurement is evaluated to be
\begin{eqnarray}
\label{parity-work}
 W&=&  W_{++}+W_{+-} +W_{-+}+W_{--}-W_{LO}=\hbar \omega\frac{\kappa^2}{4\gamma^2}\frac{2 \sigma^2_{\Delta n}}{\pi} -2\hbar \omega\beta^2\\ \nonumber
 &=&\frac{\hbar \omega}{2\pi} \frac{2\beta^2\kappa^2 (1-\kappa^2) \bar n^2}{2\beta^2 +(1-\kappa^2)(1+2\beta^2)\bar n} -2\hbar \omega\beta^2.
\end{eqnarray}
The positive part (work gain) of Eq. (\ref{parity-work}), is similar to its counterpart Eq. (52) for the work gain by fine grained homodyning, but in Eq. (\ref{parity-work}) the work is smaller by a factor of $\frac{1}{2\pi}$, since the phase is not recorded by sign measurement.

The minimum mean number of quanta for non-zero efficiency by sign measurement is $\bar n=2 \pi$ as opposed to $\bar n=1$ for fine-grained homodyne WOF.
For large $\bar n$, $W$ is optimized when $2\beta^2\approx\sqrt{\frac{\bar n}{2 \pi}}$ and $1-\kappa^2\approx\frac{1}{\sqrt{\bar n}}$, the extractable work then being 
\begin{equation}
 W\approx \frac{\hbar \omega}{2\pi}[\bar n- 2(1+\sqrt{2\pi}) \sqrt{\bar n}+1+\sqrt{2\pi}+ O(\frac{1}{\sqrt{\bar n}},\frac{1}{\bar n})].
\end{equation}

\begin{figure}
\includegraphics[width=8cm]{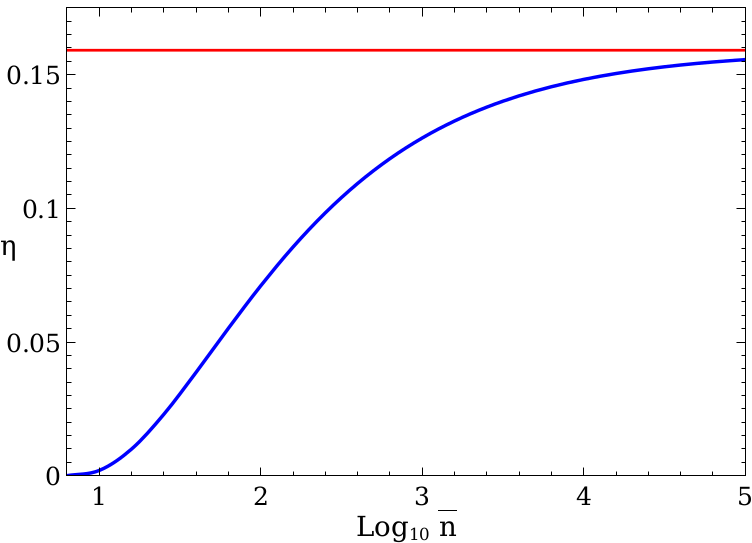}
\caption{\label{effiparity}
Efficiency $\eta=W/\hbar \omega\bar n$ plotted versus the mean number of input quanta $\bar n$ for the scheme of sign measurement WOF. The red line represent the maximal $\eta=1/2\pi$.
}
\end{figure}

The efficiency of this scheme is bounded by $\frac{1}{2 \pi}$ (see Fig.\ref{effiparity}).

The mutual information gain by sign measurement is given by (App.\ref{app:F})
\begin{eqnarray}
\label{IP}
I= -\int \int P(x,p) S(p(a,b|x,y)) dx dp+S(p(a,b)),
\end{eqnarray}
where $a,b\in \{+,-\}$. The entropy gain by the detectors for the sign measurement is  $S(p(a,b))= \ln 4$, i.e
\begin{equation}
\label{IRP}
 I_D=\ln 4.
\end{equation}

This mutual information is evaluated by taking the logarithm of the probabilities
\begin{equation}
\label{pmi-a}
 p(+,-|\alpha)=\int_0^{\infty}\int^0_{-\infty} P(\Delta n_{x},\Delta n_{p}|\alpha)d\Delta n_{x} d\Delta n_{p}= \frac{1}{4}(1+ {\rm Erf}[\frac{\mu_x}{\sqrt{2}\sigma_\alpha}])(1- {\rm Erf}[\frac{\mu_p}{\sqrt{2}\sigma_\alpha}]),
\end{equation}
where $\mu_x=\sqrt{2(1-\kappa^2)}\beta\ {\rm Re\ }\alpha$, $\mu_p=\sqrt{2(1-\kappa^2)}\beta\ {\rm Im\ }\alpha$, and $\sigma_\alpha=\left[ \frac{(1-\kappa^2)|\alpha|^2}{2}+ \beta^2 \right]^{1/2}$.
\begin{figure}
\includegraphics[width=8cm]{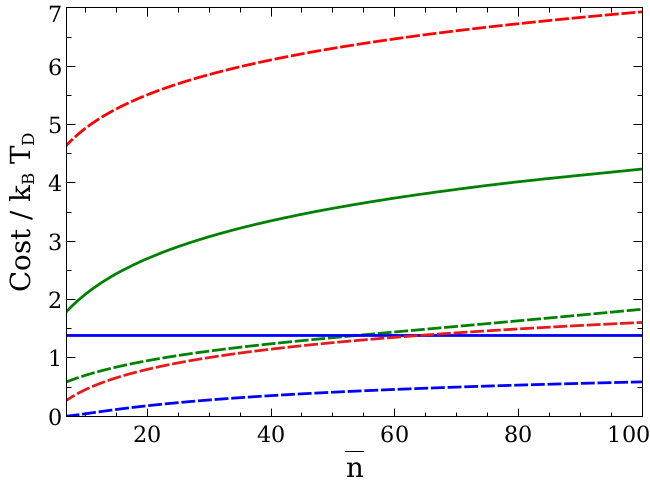}
\caption{\label{cost}
The cost of erasing the detector information (solid lines) after the completion of WOF (Eqs. (\ref{IRH},\ref{IRPC}, \ref{IRP})) and feedforward (dashed lines) (Eqs. (\ref{IH},\ref{IPC},\ref{IP})) are plotted upon normalization by $k_BT_D$, where $T_D$ is the environment temperature at which the detector is kept. The red, green and blue colors correspond to the small-fraction homodyne, photocount and sign WOF schemes, respectively. For small-fraction photocount BS transmissivity $\kappa^2=0.75$ has been considered. Clearly, $Q_D,~E_F\ll W$ for $\hbar\omega= k_BT_D.$
}
\end{figure}

\noindent Similarly,
\begin{eqnarray}
 p(+,+|\alpha)&=& \frac{1}{4}(1+ {\rm Erf}[\frac{\mu_x}{\sqrt{2}\sigma_\alpha}])(1+ {\rm Erf}[\frac{\mu_p}{\sqrt{2}\sigma_\alpha}]) \\ \nonumber
 p(-,+|\alpha)&=& \frac{1}{4}[(1- {\rm Erf}[\frac{\mu_x}{\sqrt{2}\sigma_\alpha}])(1+ {\rm Erf}[\frac{\mu_p}{\sqrt{2}\sigma_\alpha}]) \\ \nonumber
 p(-,-|\alpha)&=& \frac{1}{4}(1- {\rm Erf}[\frac{\mu_x}{\sqrt{2}\sigma_\alpha}])(1- {\rm Erf}[\frac{\mu_p}{\sqrt{2}\sigma_\alpha}]).
\end{eqnarray}
We find numerically that the lower bound of feedforward cost, $E_F\geq k_BT_DI$, is much lower compared to the fine-grained homodyne case (see Fig. \ref{cost}).

These results for work extraction from sign measurements may be compared to those of the recently proposed  Szilard/ Maxwell Demon binary measurement engines \cite{Vid2016}: A scheme where two thermal fields with $\bar n$ photons each are incident on two highly transmitting BS. A photon click or no-click is registered for the reflected part in two detectors resulting in two bits of information at most.
  If a detector clicks  with probability $1/2$, then $\bar n$ of the corresponding output  field increases to $(3/2) \bar n$. For no click, the mean decreases to $(1/2) \bar n$. Only events where one detector clicks and the other one does not (in $50\%$ of the cases) produces a net photocurrent that charges a capacitor, with $(1/2) \bar n$ photons convertible to 
photocurrent. Since the  two beams have in  total $2 \bar n$, only  $1/4$ of the input energy contributes to work, so that the efficiency bound is $1/4$. Optimization of the click probabilities yields an  efficiency  bound to  $\sim 0.3$ as compared to near-unity efficiency for $\bar n\gg 1$ by our small-fraction homodyne WOF in Sec. \ref{secIVC}


The comparison of the efficiency bound obtained by such binary methods with fine-grained WOF shows the clear superiority of the latter. In contrast, the sign measurement provides comparable performance to Maxwell-Demon binary measurement engines.

\section{Can non-selective measurements (NSM) yield work?}
\label{secVI}
\noindent Since WOF relies on selective measurements that provide mutual information on the input state, a basic question is whether NSM, which do not provide mutual information, can yield work. In Sec. \ref{secVIA} we show that NSM are indeed useless for WOF. By contrast, in Sec. \ref{secVIB} we show that NSM in a basis that does not commute with the Hamiltonian can yield not only heat (as shown in ref. \cite{Talkner1}), but also ergotropy. Finally, in Sec. \ref{secVIC} we show that NSM of correlated modes can also yield work.

\subsection{NSM of a small fraction: No work}
\label{secVIA}

\noindent Consider an arbitrary generalized positive operator valued measurement (POVM), represented by Kraus operators $K_i^\dagger K_i$ for different outcome $i$, that satisfies $\sum_iK_i^\dagger K_i= \mathcal{I}$
where $\mathcal{I}$ denotes the identity operator \cite{NielsenBOOK00}.
If we measure the reflected part (see Fig. \ref{schemewof}) and find the $i$-th outcome corresponding to the Kraus operator $K_i^\dagger K_i$, then the post-measured transmitted state is
given by
\begin{equation}
 \rho(i)=\int \! \int \frac{p(i|\alpha)  P(\alpha)}{p(i)}  |\kappa \alpha \rangle \langle \kappa \alpha | d^2 \alpha, 
\end{equation}
where $p(i|\alpha)= \mbox{Tr}[K_i^\dagger K_i |\alpha \rangle \langle \alpha |]$.
Therefore, the post-measured for NSM state is 
\begin{eqnarray}
\label{NSMinv}
 \rho(NSM)&=&\int \! \int \sum_i p(i) \frac{p(i|\alpha)  P(\alpha)}{p(i)}  |\kappa \alpha \rangle \langle \kappa \alpha | d^2 \alpha \\ \nonumber
 &=& \int \! \int P(\alpha)  |\kappa \alpha \rangle \langle \kappa \alpha | d^2 \alpha \\ \nonumber
 &=&\frac{1}{\kappa^2} \int \! \int P(\frac{\alpha}{\kappa}) |\alpha \rangle \langle \alpha | d^2 \alpha. 
\end{eqnarray}
This holds true for any complete set of measurements, as 
\begin{eqnarray}
 \sum_ip(i|\alpha)=\mbox{Tr}[\sum_i K_i^\dagger K_i |\alpha \rangle \langle \alpha |]=\mbox{Tr}[\mathcal{I} |\alpha \rangle \langle \alpha |]=1
 \end{eqnarray}
 using the linearity of the trace. From Eq. (\ref{NSMinv}), we see that the form of the input P-distribution remains unaltered for NSM, and thus the distribution remains thermal with modified mean quanta $\bar n \rightarrow \kappa^2 \bar n$.
 Therefore, NSM is a no-go strategy for WOF, where feedforward of the measurement result is essential.

\subsection{NSM in a non-commuting basis with the Hamiltonian: heat and ergotropy}
\label{secVIB}

\noindent   If we perform a NSM in a basis $\{|i\rangle\}$ which does not commute with energy basis, the state becomes diagonal in this basis 
  \begin{equation}
 \rho_{NSM}=\sum_i p_i |i\rangle \langle i|.
\end{equation}
Since this basis is off-diagonal in energy eigenbasis
  \begin{equation}
 |i\rangle =\sum_n c_{n,i}|E_n\rangle,
\end{equation}
where $|E_n\rangle$s are the energy eigenstate. Therefore, the post-measured state following a NSM in a basis that is non-commuting with $H$ ($[\rho_{NSM},H]\neq0$) is non-passive, since a passive state is always diagonal in energy basis (Sec. \ref{secII}).

In \cite{Talkner1}, the authors showed that work can be extracted from a single-temperature thermal resource and measurement without feedforward in a $4-$stroke engine by a protocol, which we here modify to account for the possible non-passivity of the post-measured state:

1) The system which is initially in equilibrium with a heat bath at temperature $T$ (and thus in a diagonal state in the energy basis $\rho_{I}(\lambda_i)= \sum_{n}p^{eq}_{n}(\lambda_{i})|E_n\rangle \langle E_n |$) undergoes an adiabatic transformation by changing its energy level spacings from $\lambda_i$ to $\lambda_f$ without changing the population. 
Work is thereby done on the system, in the amount
\begin{equation}\label{w1}
W_{I}=\sum_{n}\left [E_{n}(\lambda_{f})-E_{n}(\lambda_{i}) \right ]p_{n}^{eq}(\lambda_{i})~. 
\end{equation}

2)The system is then measured in a basis other than the energy eigenbasis: While keeping the Hamiltonian $H(\lambda_{f})$ fixed, an impulsive measurement with possible outcomes $M_j$, $j=1 \to N$, of an observable that does not commute with $H(\lambda_f)$ is performed on the system.  This state change implies a change of the occupation probabilities of the energy eigenstates: 
\begin{equation}
\rho_{I}(\lambda_f)\rightarrow\rho_{NSM}=\sum_{j}M_{j}^\dagger\rho_{I}(\lambda_{f})M_{j}\:.
\label{rpm}
\end{equation}
This measurement acts as a hot bath that imparts heat into the system in the amount
\begin{equation}
Q_M= \sum_{m,n} [E_m(\lambda_f)-E_n(\lambda_f)] T_{mn} p^{eq}_n(\lambda_i),
\label{QMH}
\end{equation}
where $T_{mn}= \sum_j|\langle E_m| M_j|E_n\rangle|^2$ denotes the transition probability from $|E_n\rangle$ to $|E_m\rangle$.
One can view the heat $Q_M$ to be provided by a hot bath at temperature $T_M$.

As opposed to ref. \cite{Talkner1}, we find that the NSM can yield not only heat but also ergotropy $\Delta \mathcal{W}_{\rm NSM}$, whose upper bound can be obtained as
\begin{equation}
\label{coh-ergo-prl}
 \Delta \mathcal{W}_{\rm NSM}\leq E(\rho_{NSM})- E(\rho_{T'}),
\end{equation}
where $\rho_{T'}$ is a thermal state with Hamiltonian $H(\lambda_{f})$ at temperature $T'$, such that $S(\rho_{NSM})=S(\rho_{T'})$.
Thus, in contrast to a four stroke engine, where a hot bath renders the system in a higher-energy but still passive state, such an NSM can change the character of the energy state distribution.

 In the second adiabatic step the parameter changes from $\lambda_f$ back to the initial value $\lambda_i$. The work done by the system is then given by
\begin{equation}\label{w2}
W_{II}=  \sum_{n}\left [E_{n}(\lambda_{i})-E_{n}(\lambda_{f}) \right ]p_{n}^{NSM},
\end{equation}
where $p_{n}^{NSM}$, the probability of finding the $n$th eigenstate in the post-measurement state (Eq. (\ref{rpm})), is given by
\begin{equation}
p_{n}^{NSM} \equiv \langle n;\lambda | \rho_{NSM}|n;\lambda\rangle.
\label{pnpm}
\end{equation}
4) The final step  is thermalization with a cold bath at temperature $T_c$.

The efficiency of this scheme in ref. \cite{Talkner1} is given by
\begin{equation}
\eta = \frac{-(W_{I}+W_{II})}{Q_M}.
\label{eta}
\end{equation}
As noted above, the treatment in ref. \cite{Talkner1} has not allowed for the possibility that the measurement may also impart ergotropy $\Delta \mathcal{W}_{\rm NSM}$ to the system, as does a non-passive (e.g. squeezed) bath \cite{niedenzu18quantum}. The appropriate efficiency bound then becomes
\begin{equation}
 \eta_{\rm max}\leq 1- \frac{T_c}{T_M}\frac{Q_M}{Q_M+\Delta \mathcal{W}_{NSM}},
\end{equation}
which can be evaluated by Eq. (\ref{coh-ergo-prl}).
This efficiency  may {\it exceed the Carnot bound}, thus proving that this machine is not a heat engine.

\subsection{NSM in a mode-correlated cycle}
\label{secVIC}
\noindent Here we consider work via NSM from two oscillator modes, hot(h) and cold(c), that are correlated by their interaction, unlike the input modes in Sec. \ref{secIV}.
Let us consider a brief QND measurement that decorrelates modes, thus altering their correlation energy. 
Subsequent periodic modulation of the modes frequencies allows for work extraction following {\it an impulsive} measurement by a detector D. 
The total Hamiltonian describing a system consisting of 2 (hot-h and cold-c) interacting modes described by the Hamiltonian $H_S=H_h+H_c+H_{hc}$ and a detector is
\begin{equation}
\label{12.a110}
 H_{\rm tot}=H_{\rm S}+H_{\rm SD}
\end{equation}
where $H_{\rm SD}$ is the  impulsive system-detector interaction that does not commute with $H_{\rm S}$ ($[H_{\rm S},H_{\rm SD}]\neq0$).
This total Hamiltonian is assumed to be $\tau$-periodic, $H_{\rm tot}(\tau)=H_{\rm tot}(0)$.
Work extraction comes about because the NSM changes the intermode mean correlation energy $\langle H_{hc}\rangle$.

When the detector is traced out, the entropy and energy of the single-mode change since the NSM decorrelates the modes, thereby increasing their correlation energy by
\begin{equation}
 \Delta E_{\rm D}=-\langle H_{ hc}\rangle_{\rm Eq} >0.
\end{equation}
This scenario stands in contrast to Landauer's \cite{LandauerIBM61}, where such correlations are not accounted for. 
If the cycle duration is shorter than the correlation time, $t_{\rm cycle}<t_c$, but longer than the time needed to perform the measurement, the maximal amount of extractable work, without measurement readout (for an NSM) is given by
\begin{equation}
(W_{\rm NSM})_{\rm max}=
\Delta E_{\rm D}-T_D\Delta {\cal S}_{\rm D}, 
\end{equation}
where $\Delta{\cal S}_{\rm D}$ is the entropy\index{entropy} increase of the detector due to the NSM.
 
The energy $\Delta E_{\rm D}$ consumed by the detector can be a thermal noisy pulse, described by a passive state, so that neither the detector nor these modes can store ergotropy.
The NSM-based cycle converts such passive input into a non-passive output state capable of delivering work.

Such a cycle exemplifies the conclusion that, upon entangling the initially uncorrelated passive (but non-thermal) states of distinct subsystems, here the intermode and the detector, the state of one subsystem (here the hot mode) may become non-passive and thus deliver work.

The maximum work (per cycle) extractable from a selective measurement, $(W_{\rm sel})_{\rm max}$, clearly exceeds the NSM-work, $(W_{\rm NSM})_{\rm max}$: 
\begin{equation}
 (W_{\rm sel})_{\rm max}=
(W_{\rm NSM})_{\rm max} +W,
\end{equation}
 where $W$ is the work obtained by WOF in Sec. \ref{secIV} or \ref{secV} in the absence of $\Delta E_D$.
The extra work $(W_{\rm NSM})_{\rm max}$ stems from  correlations or entanglement unaccounted for by the Landauer principle. 

Remarkably, an NSM in this scenario allows for work extraction from a bath at $T_D=0$, without information gain:
The reason is that the correlation energy is always negative, even at $T_D=0$.
Hence, decorrelation of the modes through a measurement increases the total energy allowing the cycle to be triggered, yielding the extractable work
\begin{equation}
 (W_{\rm sel})_{\rm max}=
(W_{\rm NSM})_{\rm max}> 0.
\end{equation}

A similar situation arises for a system and a bath that adhere to the spin-boson model \cite{GelbwaserPRA13}, where work extraction via NSM can only take place within the correlation time scales.
The joint,  entangled multimode state initially at equilibrium, $\rho_{\rm Eq}$, is changed \cite{Erez2008} to a product state by the impulsive NSM \cite{GelbwaserPRA13,Erez2008}.

\section{Conclusions}
\label{secVII}

\noindent  Our  comparative analysis of heat to work conversion in few-mode setups  by  either unitary (reversible) manipulations  or measurements has led to the following findings:

    A.  The drawbacks of reversible manipulations have been shown (Sec. \ref{secIII}) to be  (i) the need for adiabaticity in order  to achieve high efficiency of work extraction, resulting in vanishing power; and (ii) the practical difficulty to manipulate the mode frequencies.  These drawbacks are partly circumvented by measurement-based schemes, where power is mainly limited
    by the feedforward and detector-resetting time, and is {\it independent of efficiency}.
 
    B.  As compared to the previously proposed work extraction by measuring a variable of the entire input \cite{Vid2016}, we have shown  (Sec. \ref{secIV}) that it is advantageous to measure only a small fraction of the input and extract work from the dominant, unmeasured fraction , by generalizing   our recently proposed method of work by observation and feedforward (WOF) \cite{WOF}.  The main advantage of measuring  a small fraction, either by photocount or by homodyning,  is that it bears much smaller cost in terms of information (entropy) consumed by feedforward and by resetting the detectors (after WOF has been completed). 
    
    C.  We  have argued (Sec. \ref{secIV-1}) that, practically, the resetting  of the detectors  should preferably be done as fast as possible, since detector cooling to its initial temperature may carry a modest energy and entropy cost compared to the extracted work. 
    
    D. Measurements with partial resolution (coarse graining)   have been shown  (Sec. \ref{secV}) to yield much less information as well as work and efficiency than their fine-graining counterparts, thereby establishing the rapport of work and information extraction. Yet, WOF based on extreme coarse-graining of a small fraction  has been shown to favorably compare with binary-measurement  (Maxwell-demon) information machines \cite{Vid2016,SagawaPRL08,Elouard17,Elouard18}.
    
    E. Finally,  unread or non-selective measurements (NSM) \cite{Kurizkibook} have been  shown (Sec. \ref{secVI}) to yield no work when applied in WOF. Yet, they may extract work when performed in a basis that does not commute with the Hamiltonian: In fact, we have shown that NSM may yield considerably more work than previously proposed \cite{Talkner1,Maron}.  In scenarios where the modes are nonlinearly correlated, NSM  has been noted to yield  work from the intermode correlation energy, a consideration absent in Landauer’s principle \cite{LandauerIBM61}.  These scenarios are analogous  to  work extraction by NSM   from system-bath correlations in the non-Markovian time-domain \cite{GelbwaserPRA13}.
    
The present analysis has not only conceptual but also practical merit, in particular for optical setups and their acoustic counterparts. While the {\it spatial profile} of electromagnetic or acoustic field propagation and its mode decomposition are well controlled  by simple elements ( collimators, beam splitters, lenses etc.), {\it temporal fluctuations} are much harder to control. Our comparative analysis has presented guidelines to the alternative methods by which such control can be accomplished for single-mode, i.e. spatially well collimated propagation of  thermal noise, resulting in optimized work extraction.  The bounds on  this work extraction and the corresponding power have been quantified by the minimal costs required for these tasks, i.e. information transfer for feedforward and detector resetting.

These bounds are important for determining the feasibility of few-quanta conversion from heat to work. Optical elements have been shown to allow the increased concentration of sunlight so that  the stationary power that arrives at the detector on average is multiphoton, but it has thus far been unclear  what level of power suffices for work generation. Our analysis  makes us cautiously optimistic that  this task may be experimentally accomplished with a few photons.  It may manifest itself, e.g., as the transformation of concentrated sunlight input into nearly-coherent or number-squeezed light at the output and thereby produce reduced quantum fluctuations in an optomechanical device \cite{Gelbwaser_2015_b}. Alternatively, thermal light input may yield low-noise (low-entropy) photocurrent \cite{Dong2021,GKphotocurrent} that can be instrumental for quantum operation of electronic devices. 

\acknowledgments
\noindent A.M. thanks Arnab Chakrabarti, Nilakantha Meher and Saikat Sur of WIS for useful discussions.
T.O. is supported by the Czech Science Foundation, Grant No. 20-27994S.  G.K. is supported by ISF, DFG (FOR 2724), QUANTERA (PACE-IN) and NSF-BSF.

\appendix
\section{\label{AppendixA} Reversible work extraction}
\label{app:A}
The entropy of the thermal mode with the mean quanta $\bar n$ is
\begin{eqnarray}
S (\bar n) = k_B[(\bar n + 1)\ln (\bar n + 1) - \bar n\ln \bar n] ,
\end{eqnarray}
and the temperature is
\begin{eqnarray}
k_BT=\frac{\hbar \omega}{\ln \left(1+\frac{1}{\bar n} \right)}.
\end{eqnarray}
Thus, the initial total entropy is 
\begin{eqnarray}
S_0 = k_B[(\bar n + 1)\ln (\bar n + 1) - \bar n\ln \bar n + (N-1)( (\bar n_c + 1)\ln (\bar n_c + 1) - \bar n_c\ln \bar n_c)],
\end{eqnarray}
and the initial mean energy is
\begin{eqnarray}
E_0 = \hbar \omega[\bar n + (N-1)\bar n_c] .
\end{eqnarray}
Since each mode has now $\bar n_f$ quanta on average, the total entropy is
\begin{eqnarray}
S_f  &=& N k_B\left[ (\bar n_f + 1)\ln (\bar n_f + 1) - \bar n_f\ln \bar n_f  \right], 
\end{eqnarray}
and the mean energy is
\begin{eqnarray}
E_f = N \hbar \omega \bar n_f.
\end{eqnarray}

In the classical limit $\bar n \gg 1$, 
 one can approximate $T\propto \bar n$, and
\begin{eqnarray}
S_0 (\bar n) &\approx& k_B [\ln \bar n + (N-1) \ln \bar n_c+N] , \\
S_f (\bar n_f)  &\approx& Nk_B[1+\ln \bar n_f].
\end{eqnarray}
\section{\label{AppendixB}Thermodynamics of a single oscillator mode}
\label{app:B}
Starting from the partition function
\begin{eqnarray}
Z=\sum_{n=0}^{\infty}\exp \left(-\frac{\hbar \omega}{k_BT}n \right) = \frac{1}{1-\exp \left(-\frac{\hbar \omega}{k_BT} \right)}
=  \frac{1}{1-\exp \left(-\beta \hbar \omega \right)},
\end{eqnarray}
one finds the mean energy
\begin{eqnarray}
E = \frac{1}{Z}\frac{\partial Z}{\partial (1/k_B T)} = \frac{\hbar \omega}{\exp \left( \frac{\hbar \omega}{kT}\right)-1} = \hbar \omega \bar n .
\end{eqnarray}
Expressing the relationship between temperature and mean photon number as
\begin{eqnarray}
T=\frac{\hbar \omega}{k_B \ln \left( 1+\frac{1}{\bar n} \right)}
\label{EqTempT}
\end{eqnarray}
we can express the partition function as
\begin{eqnarray}
Z=\bar n + 1,
\end{eqnarray}
entropy as
\begin{eqnarray}
S = k_B\left( \ln Z +  E/k_BT \right) = k_B\left[ (\bar n + 1)\ln (\bar n + 1) - \bar n  \ln \bar n \right],
\label{EqEntropyS}
\end{eqnarray}
and free energy
\begin{eqnarray}
\mathcal{F}=-k_BT \ln Z = -\hbar \omega + k_BT \ln \left[\exp \left(\frac{\hbar \omega}{k_BT} \right)-1 \right]
=-\hbar \omega \frac{\ln (1+\bar n)}{\ln \left(1+\frac{1}{\bar n} \right)}.
\end{eqnarray}
We can write the first law (or, more precisely, the combined theorem) of thermodynamics as
\begin{eqnarray}
dE = TdS + {\cal P}d\omega ,
\end{eqnarray}
where $TdS = \hbar \omega d\bar n$ is the heat entering the system and ${\cal P}d\omega$ is the work done on the system by changing the frequency, where the ``pressure'' ${\cal P}$ is given by the derivative $\mathcal{F}$ with respect to $\omega$
\begin{eqnarray}
{\cal P} = \left(\frac{\partial \mathcal{F}}{\partial \omega} \right)_{T} = \frac{\hbar}{\exp \left(\frac{\hbar \omega}{kT} \right)-1} = \hbar \bar n .
\end{eqnarray}
Thus, we can express the work done on the system as free energy change, and heat entering the system as entropy change during an isothermal process
\begin{eqnarray}
W=\mathcal{F}_2-\mathcal{F}_1 &=& \hbar (\omega_1-\omega_2) + kT \ln \frac{\exp \left(\frac{\hbar \omega_2}{kT} \right)-1 }{\exp \left(\frac{\hbar \omega_1}{kT} \right)-1 }, \\ \nonumber
&=& 
\hbar \left[\omega_1 \frac{\ln (1+\bar n_1)}{\ln \left(1+\frac{1}{\bar n_1} \right)}
- \omega_2 \frac{\ln (1+\bar n_2)}{\ln \left(1+\frac{1}{\bar n_2} \right)} \right]  \\
Q=T(S_2-S_1) &=& \hbar \left[ \frac{\omega_2}{\exp\left(\frac{\hbar \omega_2}{kT} \right)-1 } - \frac{\omega_1}{\exp\left(\frac{\hbar \omega_1}{kT} \right)-1 } \right] + kT \ln \frac{1-\exp \left(-\frac{\hbar \omega_1}{kT} \right)}{1-\exp \left(-\frac{\hbar \omega_2}{kT} \right)}
\nonumber \\
&=& \hbar \omega_2 \frac{(\bar n_2 + 1)\ln (\bar n_2 + 1) - \bar n_2  \ln \bar n_2}{\ln \left(1+\frac{1}{\bar n_2} \right)}
- \hbar \omega_1 \frac{(\bar n_1 + 1)\ln (\bar n_1 + 1) - \bar n_1  \ln \bar n_1}{\ln \left(1+\frac{1}{\bar n_1} \right)} .
\end{eqnarray}

Considering the limit $\hbar \omega_2 \gg k_BT$ one can find the work necessary to isothermally compress the oscillator to infinite $\omega$, as well as the corresponding heat (using here $\bar n$ and $\omega$ instead of $\bar n_1$ and $\omega_1$),
\begin{eqnarray}
\label{EqWInfty}
W_{\infty} = -\mathcal{F} &=& \hbar \omega - k_BT \ln \left[\exp \left(\frac{\hbar \omega}{k_BT} \right) -1\right] \nonumber \\
&=& \hbar \omega \frac{\ln (1+ \bar n)}{\ln \left(1+\frac{1}{\bar n} \right)} , \\
Q_{\infty}=-TS &=& -\hbar \omega \left[\bar n + \frac{\ln \left( 1+\bar n \right)}{\ln \left( 1+\frac{1}{\bar n}\right)} \right] .
\label{EqQInf}
\end{eqnarray}
As can be seen, $W_{\infty}+Q_{\infty}=-\hbar \omega\bar n$, i.e., during an isothermal process the work spent on increasing $\omega$ plus the initial energy $\hbar \omega\bar n$ are converted into heat going to the environment. Note that in the limit $k_BT\gg \hbar \omega$, or $\bar n \gg 1$, one gets
\begin{eqnarray}
\label{EqWInfty2}
W_{\infty} &\approx& \hbar \omega \left[\left( \bar  n+\frac{1}{2}\right) \ln \bar n + 1 \right] , \\
Q_{\infty}&\approx& -\hbar \omega \left[\left( \bar  n+\frac{1}{2}\right) \ln \bar n + \bar n + 1 \right].
\end{eqnarray}

\section{\label{AppendixC} Photocount of a reflected thermal beam}
\label{app:C}
When a Fock state $|n\rangle$ is incident on a beam-splitter (BS) with transmissivity $\kappa^2$,
the transmitted state \cite{Scully_BOOK} is 
\begin{equation}
 |n,0\rangle_{\rm out}= \sum_{q=0}^n \sqrt{\frac{n!}{(n-q)!q!}} (\kappa)^q(\sqrt{1-\kappa^2})^{n-q}|q,n-q\rangle.
\end{equation}
If we  detect $m$ photons in the reflected beam,  the resulting transmitted state is $|n-m\rangle$. This event has the probability
\begin{equation}
p'_m=\frac{n!}{(n-m)!m!}(\kappa^2)^{n-m}(1-\kappa^2)^{m} 
\end{equation}
For a thermal input as in Eq. (\ref{fockthermal}),  detecting $m$  quanta in the reflected  beam  has the  probability
\begin{equation}
\label{PN}
 p_m= \frac{(1-\kappa^2)^m (1-e^{-\frac{\hbar \omega }{k_B T}})e^{-\frac{\hbar \omega }{k_B T} m}}{(1-e^{-\frac{\hbar \omega }{k_B T}}\kappa^2)^{(m+1)}}.
\end{equation}
As the BS does not change the distribution of the input, the reflected beam corresponds to a thermal distribution with mean quanta number  $(1-\kappa^2)\bar n$ (Eq. \ref{effective-T}).
The average energy of the post-measured state is 
\begin{equation}
 E_m=\hbar \omega \bar n_m=\hbar \omega \sum_{n=0}^\infty p(n|m) n=\hbar \omega \frac{(1+m)\kappa^2}{e^{\frac{\hbar \omega }{k_B T}} -\kappa^2}.
\end{equation}
Assume a thermal state with mean photon number $\bar n$ entering a beam splitter with reflectivity
\begin{equation}
 R=1-\kappa^2.
\end{equation}
 The reduced density matrix for the reflected corresponds to a thermal state with mean photon number $R\bar n$. Let us assume that $m$ photons in the reflected beam were detected. The conditional probability distribution of the photon number $n$ is evaluated
\begin{eqnarray}
 p(n|m) = \frac{p(n \wedge m)}{p_{\rm refl}(m)},
\end{eqnarray}
where $p(n \wedge m)$ denotes the joint probability of having $m$ photons in the reflected beam and $n$ photons in the transmitted beam. It is given by
\begin{eqnarray}
p(n \wedge m) &=& p(n \wedge m| n+m) p_{\rm in}(n+m), \\
p(n \wedge m| n+m) &=& {n+m \choose m}R^m (1-R)^n , \\
p_{\rm in}(n+m) &=& \frac{\bar n^{n+m}}{(\bar n +1)^{n+m+1}}, \\
p_{\rm refl}(m) &=& \frac{(R\bar n)^{m}}{(R\bar n+1)^{m+1}}, 
\end{eqnarray}
 where $p_{R}(m)$ is the marginal probability of having $m$ photons in the reflected beam, and $p_{\rm in}(n+m)$ is the probability of having $n+m$ photons in the incoming beam.
Using these equations one finds 
\begin{eqnarray}
p(n|m) &=&  {n+m \choose m}R^m (1-R)^n \frac{\bar n^{n+m}}{(\bar n +1)^{n+m+1}}
\frac{(R\bar n+1)^{m+1}}{(R\bar n)^{m}} \nonumber \\
&=& \frac{(n+m)!}{n!m!}(1-R)^n \frac{\bar n^{n} (R\bar n+1)^{m+1}}{(\bar n +1)^{n+m+1}} .
\label{EqExact}
\end{eqnarray} 
This result is exact. If the numbers $n,m$ are too large so that computation of the factorials is impractical, one can use an approximation based on the Stirling formula
\begin{eqnarray}
n! \approx \sqrt{2\pi n}\left(\frac{n}{e} \right)^n
\end{eqnarray}
to get
\begin{eqnarray}
p(n|m) &\approx& \sqrt{\frac{n+m}{2\pi nm}}\left(\frac{n}{m} \right)^m \left(1+\frac{m}{n} \right)^{n+m}
 (1-R)^n \frac{\bar n^{n} (R\bar n+1)^{m+1}}{(\bar n +1)^{n+m+1}} .
\label{EqApprox}
\end{eqnarray}
\begin{figure}
\includegraphics[width=8cm]{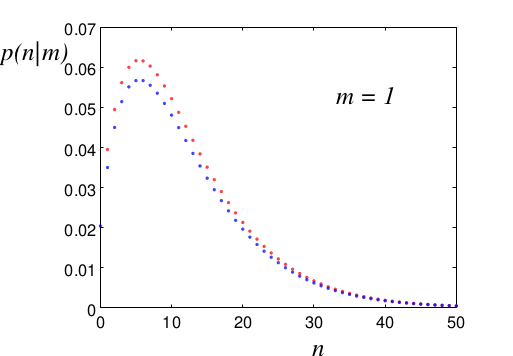}
\includegraphics[width=8cm]{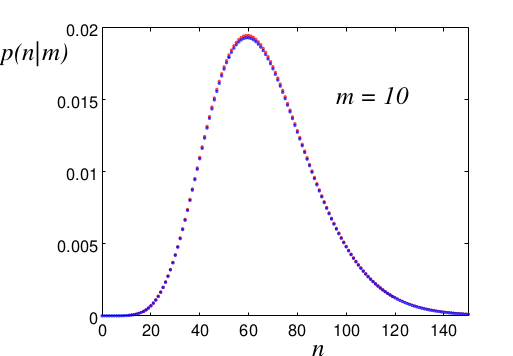}
\caption{\label{f-m}
Comparison of the exact conditional probability distribution $p(n|m)$ as in Eq. (\ref{EqExact}) (blue) with the approximate formula of Eq. (\ref{EqApprox}) (red).
}
\end{figure}

\section{\label{AppendixD} Phase-plane distribution of the post-measured state following small-fraction photocount}
\label{app:D}
The distribution of $\alpha$, conditioned on the detection of quanta number   $m$  is
\begin{eqnarray}
 P(\alpha |m) = \frac{p(m|\alpha) P(\alpha) }{p(m)}.
\label{Eqa}
\end{eqnarray}
The unmeasured (transmitted) field mode has the state (conditional on the detection of  $m$)
\begin{eqnarray}
\hat \varrho (n) 
&=& \frac{1}{\kappa^2} \int \! \int P\left(\frac{\alpha}{\kappa}|m\right) |\alpha \rangle \langle \alpha | d^2 \alpha.
\label{eqb}
\end{eqnarray}
In small-fraction photocunt, the distribution of detected photons for a coherent state input $|\alpha\rangle$ yields a Poissonian statistics
with the mean number of quanta $\lambda= (1-\kappa^2)|\alpha|^2$,
\begin{equation}
 p(m|\alpha)=e^{-\lambda}\frac{\lambda^m}{m!}.
\end{equation}
In Eq. (\ref{Eqa}) $p(m)$ is the quanta number distribution of a thermal state with mean quanta number $(1-\kappa^2)\bar n$, obtained according to Eq. (\ref{effective-T}).

%
\section{\label{AppendixE} Work optimization for phase-sensitive measurement}
\label{app:E}

Upon substituting $\xi= 2\beta^2$, and $\epsilon = 1-\kappa^2$, the work in Eq. (\ref{wdis}) is optimized for 
\begin{eqnarray}
\xi=\frac{\sqrt{\bar n (1-\epsilon)}-1}{1+\frac{1}{\epsilon \bar n}}.
\label{Apxi}
\end{eqnarray}
and
\begin{eqnarray}
\epsilon = \frac{\sqrt{\bar n -\sqrt{\bar n}+1}-1}{\bar n}.
\label{Apepsilon}
\end{eqnarray}
Using these values one gets the maximal extractable work in Eq. (\ref{Woptimum}) as
\begin{eqnarray}
W_{\rm max}/\hbar \omega \approx
\left(\sqrt{\bar n - \sqrt{\bar n}+1}-1 \right)^2  \left( 1-\frac{1}{\sqrt{\bar n}} \right).
\label{WoptimumApp}
\end{eqnarray}

Let us optimize the extractable work in Eq. (90). substituting $\xi= 2\beta^2$ and $\epsilon = 1-\kappa^2$
one can write Eq. (90) as
\begin{eqnarray}
W/\hbar \omega = \frac{\bar n}{2\pi} \frac{1-\epsilon}{1+\frac{1}{\xi}+\frac{1}{\epsilon \bar n}}-\xi.
\label{WnAp}
\end{eqnarray}
Equating $\frac{\partial W}{\partial \xi}=0$, we get a quadratic equation
\begin{equation}
 \left(1+ \frac{1}{\epsilon \bar n}\right)^2 \xi^2 + 2 \left(1+ \frac{1}{\epsilon \bar n}\right) \xi
+ 1-\bar n (1-\epsilon)/2 \pi = 0,
\end{equation}
whose only positive root is given by
\begin{eqnarray}
\xi=\frac{\sqrt{\bar n (1-\epsilon)/2\pi}-1}{1+\frac{1}{\epsilon \bar n}}.
\label{Apxi}
\end{eqnarray}
substituting this in Eq. (\ref{WnAp}), we get
\begin{eqnarray}
W/\hbar \omega= \frac{\bar n \epsilon}{\bar n \epsilon +1} \left[\frac{\bar n}{2\pi}(1-\epsilon)-2\sqrt{\frac{\bar n}{2\pi}(1-\epsilon)}+1\right].
\end{eqnarray}
For high transmittance BS using the approximation $\sqrt{1-\epsilon}\approx 1-\epsilon/2$, we get
\begin{eqnarray}
W \approx \frac{\bar n \epsilon}{\bar n \epsilon +1}\left[(\sqrt{\bar n/2\pi}-\bar n/2\pi)y+(\sqrt{\bar n/2\pi}-1)^2\right].
\label{WnAp2}
\end{eqnarray}
Again equating $\frac{\partial W}{\partial \epsilon}=0$, we get
\begin{eqnarray}
\bar n \epsilon^2 + 2 \epsilon -1 +\sqrt{2\pi/\bar n} =0,
\end{eqnarray}
 The above Eq. has only one positive root given by
\begin{eqnarray}
\epsilon = \frac{\sqrt{\bar n -\sqrt{ 2\pi \bar n}+1}-1}{\bar n}.
\label{Apepsilon}
\end{eqnarray}

\section{\label{AppendixF} Mutual information in photocount, homodyne and sign WOF}
\label{app:F}

Say we have the input in a particular state $n$ with and we get measurement outcome $m$ with probability $p(m)$, the pointwise mutual information
\begin{equation}
 I_{\rm mn}= \ln p(m| n)-\ln p(m),
\end{equation}
quantifies the uncertainty reduced on average in the measurement outcome $m$ when the input is in state $n$ \cite{NielsenBOOK00}.
Here $p(m|n)$ is the conditional probability. If we average this pointwise mutual information over the joint probability distribution $p(m,n)$ we get the total mutual information which quantifies the correlation between the measured system and the outcomes \cite{NielsenBOOK00}, i.e.
\begin{equation}
 I= \sum_{m,n} p(m,n)I_{\rm mn}.
\end{equation}
Here we have 
\begin{equation}
\label{use1}
 \sum_n p(m,n)=p(m);~~\sum_m p(m,n)=p(n);~~ p(m,n)=p(m| n)p(n)
\end{equation}
and the Bayes' theorem reads as 
\begin{equation}
\label{use2}
 p(m| n)=\frac{p(n| m)p(m)}{p(n)}.
\end{equation}
Using Eq. (\ref{use1}) and (\ref{use2}), we get
\begin{equation}
\label{formulaI}
 I=-\sum_n p(n)S(p(m|n))+S(p(m)),
\end{equation}
where  the Shanon entropy $S(p(i))$ associated with the probability distribution $p(i)$
is given as
\begin{equation}
 S(p(i))= -\sum_i p(i)\ln p(i).
\end{equation}
We have used Eq. (\ref{formulaI}) for computing mutual information for the photocount WOF and sign measurement WOF. The sum is replaced by integral  where the continuum limit is applicable. 
We have considered natural logarithm instead of $\log_2$ in computing mutual information or entropy. However, as we are interested in calculating erasing lower bound on the cost of the detector and feedforward cost which are  $k_BT_D \ln2$ times the entropy and mutual information in bits (i.e. with $\log_2$), we compute $I$ and $I_D$ in natural logarithm units and multiply them by $k_BT_D$.

For sign measurement WOF, $P(\Delta n_{x},\Delta n_{p}|\alpha)$ in Eq. (\ref{pmi-a}) can be approximated for large quanta number as \cite{WOF}
\begin{eqnarray}
 P(\Delta n_{x},\Delta n_{p}|\alpha) \approx \frac{1}{2\pi \left[ \frac{(1-\kappa^2)|\alpha|^2}{2}+ \beta^2 \right]}
\exp \left[ {-\frac{\left(\Delta n_{x} - \sqrt{2(1-\kappa^2)}\beta\ {\rm Re\ }\alpha\right)^2
+ \left(\Delta n_{p} - \sqrt{2(1-\kappa^2)}\beta\ {\rm Im\ }\alpha\right)^2}{(1-\kappa^2)|\alpha|^2+2\beta^2}} \right].
\label{EqPnalpha}
\end{eqnarray}

For calculating mutual information for the homodyne WOF we have additionally considered properties of mutual information of two Gaussian distribution as detailed below.
The mean mutual information generated in the detection process is
\begin{eqnarray}
I = \sum_{\Delta n_x} \sum_{\Delta n_p} \int \int \ln \frac{P(x,p |\Delta n_x, \Delta n_p)}
{P(x,p)} P(x,p |\Delta n_x, \Delta n_p) P(\Delta n_x, \Delta n_p) dx dp .
\label{EqP1}
\end{eqnarray}
We have
\begin{eqnarray}
I &\approx& \langle I_x \rangle + \langle I_p \rangle,
\end{eqnarray}
where
\begin{eqnarray}
\langle I_x \rangle &\approx&  \int \int \ln \frac{P(x |\Delta n_x)}
{P(x)} P(x, |\Delta n_x)P(\Delta n_x) d\Delta n_x  dx , \\
\langle I_p \rangle &\approx&  \int \int \ln \frac{P(p |\Delta n_p)}
{P(p)} P(p, |\Delta n_p) P(\Delta n_p) d\Delta n_p  dp .
\end{eqnarray}
Here
\begin{eqnarray}
P(x,p |\Delta n_{x},\Delta n_{p}) 
\approx \frac{1}{2\pi \sigma_x^2} 
\exp \left[-\frac{(x-\bar x_{\Delta nx})^2 + (p-\bar p_{\Delta np})^2}{2\sigma_x^2} \right],
\end{eqnarray}
with
\begin{eqnarray}
\label{ApproxBarx}
\bar x_{\Delta nx} &=& \frac{ \Delta n_x}{\beta \sqrt{1-\kappa^2}\left[1+\frac{1}{\bar n (1-\kappa^2)}+\frac{1}{2\beta^2} \right]} , \\
\label{ApproxBarp}
\bar p_{\Delta np} &=& \frac{ \Delta n_p}{\beta \sqrt{1-\eta^2}\left[1+\frac{1}{\bar n (1-\kappa^2)}+\frac{1}{2\beta^2} \right]} , \\
 \sigma_x^2 &=& \frac{\bar n}{1+\frac{2\beta^2\bar n (1-\kappa^2)}{2\beta^2+\bar n (1-\kappa^2)}}.
\label{SigmaX}
\end{eqnarray}

Eq. (\ref{EqP1}) can be evaluated using the following property of a Gaussian distributions of variables $X$ and $Y$ that the mutual information is given by
\begin{eqnarray}
\langle I(X;Y)\rangle  = -\frac{1}{2} \ln \left( 1-\frac{{\rm Var}_{X,Y}^2}{{\rm Var}_X {\rm Var}_Y} \right).
\end{eqnarray}
We find 
\begin{equation}
 {\rm Var}_{x,\Delta n_x} = {\rm Var}_{p,\Delta n_p} = \epsilon \sigma_{\Delta n}^2
\end{equation}
and 
\begin{equation}
 {\rm Var}_{x} = {\rm Var}_{p} = \bar n 
\end{equation}
\begin{equation}
 {\rm Var}_{\Delta n_x} = {\rm Var}_{\Delta n_p} =\sigma_{\Delta n}^2
\end{equation}
\begin{equation}
\sigma_{\Delta n}^2 = \beta^2 + \bar n (1-\kappa^2)\left( \beta^2 + \frac{1}{2} \right) 
\end{equation}

\bibliography{wof}

\end{document}